\documentclass{article}
\usepackage{fullpage} 
\usepackage{amsmath}
\usepackage{bm}
\usepackage{graphicx}
\usepackage{multirow}
\usepackage{xfrac}
\usepackage{verbatim}
\usepackage{float}
\usepackage{color,soul}
\usepackage{lscape}
\usepackage{pdflscape}
\usepackage{authblk}
\usepackage{chngcntr}

\usepackage{titlesec}
\newcommand{\keywords}[1]{\textbf{\textit{Keywords---}} #1}

\begin{document}

\title{Two-sample aggregate data meta-analysis of medians}
\author[1]{Sean McGrath}
\author[2]{Hojoon Sohn}
\author[1]{Russell Steele}
\author[3 4, 5]{Andrea Benedetti}

\affil[1]{\small Department of Mathematics and Statistics, McGill University, Quebec, Canada}
\affil[2]{\small Department of Epidemiology, Bloomberg School of Public Health, Johns Hopkins University, Baltimore, MD, USA}
\affil[3]{\small Respiratory Epidemiology and Clinical Research Unit, McGill University Health Centre, Quebec, Canada}
\affil[4]{\small Department of Epidemiology, Biostatistics and Occupational Health, McGill University, Quebec, Canada}
\affil[5]{\small Department of Medicine, McGill University, Quebec, Canada}

\date{}
\maketitle

\begin{abstract}
We consider the problem of meta-analyzing two-group studies that report the median of the outcome. Often, these studies are excluded from meta-analysis because there are no well-established statistical methods to pool the difference of medians. To include these studies in meta-analysis, several authors have recently proposed methods to estimate the sample mean and standard deviation from the median, sample size, and several commonly reported measures of spread. Researchers frequently apply these methods to estimate the difference of means and its variance for each primary study and pool the difference of means using inverse variance weighting. In this work, we develop several methods to directly meta-analyze the difference of medians. We conduct a simulation study evaluating the performance of the proposed median-based methods and the competing transformation-based methods. The simulation results show that the median-based methods outperform the transformation-based methods when meta-analyzing studies that report the median of the outcome, especially when the outcome is skewed.  Moreover, we illustrate the various methods on a real-life data set. 
\end{abstract}


\section{Introduction}

Meta-analysis is a statistical method for combining data from several scientific studies addressing a common research question. Researchers conduct meta-analyses to obtain more precise statistical estimates of the outcome of interest, investigate sources of heterogeneity, and generalize conclusions across studies. Over the past few decades, meta-analysis has gained considerable popularity in research synthesis and is widely considered to be the gold-standard of evidence for systematic review.

Researchers often meta-analyze two-group studies that evaluate the change in an outcome across two groups. One such study design includes clinical trials with a treatment and control group. Frequently, only summary measures of the outcome variable in each group are available from the primary studies, which typically include a point estimate of the outcome and its variance. Data analysts must then decide on the effect measure to use in the analysis, which quantifies the change of the outcome between the two groups. Standard meta-analytic methods estimate a pooled effect measure by a weighted mean of the study-specific effect measures, where the weight assigned to each study are inversely proportional to the variance of the point estimate. These methods are referred to as inverse variance approaches \cite{Borenstein2010}. 

In this paper, we consider the case where the outcome variable is continuous and is measured on a scale that is meaningful and consistent across studies. In this context, two-group studies usually report the sample mean and standard deviation of the outcome in each group, and researchers typically use the difference of means as the effect measure of the primary studies. Then, researchers meta-analyze the difference of means in the two groups. Meta-analytic methods for pooling the difference of means are well-established in the literature \cite{Deeks2008}.

However, when the distribution of the outcome variable is skewed, authors often report the sample median of the outcome to better represent the center of the data. In this case, to describe the spread of the data, authors frequently report the first and third quartiles, the minimum and maximum values, or both sets of measures (e.g., see \cite{qin2016delays, hojoon2016improving}). Although the difference of medians is the most natural effect measure to use in this context for two-group studies, there are no available statistical methods to meta-analyze the difference of medians. The challenge of applying the inverse variance approach to pool the difference of medians is that the variance of the difference of medians is not reported by the primary studies. Furthermore, this quantity is difficult to estimate based on the summary measures provided by a primary study because it depends on the underlying distribution. 

Consequently, studies that report the median of the outcome are either left out of meta-analyses (e.g., see \cite{Hagiwara2014}) or researchers transform the outcome so that well-established meta-analytic methods can be applied. More specifically, several authors have recently proposed methods estimate the sample mean and standard deviation from the median, sample size, and several commonly reported measures of spread \cite{wan2014estimating, hozo2005estimating, bland2014estimating, kwon2015simulation, kwon2016, luo2016optimally}. These methods are used to estimate the difference of means and its variance for each primary study that reports medians. Then, data analysts pool the estimated difference of means using the inverse variance method.  Although these methods are widely used in practice, they suffer several limitations. First, in this analysis, one uses an estimate for the study-specific effect measure (i.e., the estimated difference of means) rather than one that can be directly calculated (i.e., the difference of medians), which introduces an additional source of error in the analysis. Moreover, the performance of these methods for estimating the sample mean and standard deviation has been shown to be heavily influenced by the skewness of the outcome distribution \cite{kwon2015simulation}, which is likely to be highly skewed when authors report medians. Finally, regardless of the accuracy of these methods for estimating sample mean and standard deviation, sample medians may better represent the center of the data for skewed outcomes and therefore may be the preferred effect measure in meta-analysis of skewed continuous outcomes.  

In this work, we develop methods to directly meta-analyze the difference of medians. We first consider several simple methods that are not based on the classical inverse variance approach. Specifically, we extend the methods proposed by McGrath et al \cite{mcgrath2018} for meta-analyzing one-group studies that report the median to the two-sample context, and we consider applying linear quantile mixed models. In order to weight studies optimally, we develop methods to estimate the variance of the median in order to pool the difference of medians using the inverse variance approach. 

We describe the existing the proposed methods to meta-analyze two-group studies that report medians in Section \ref{Methods}. In Section \ref{Simulation}, we present the results of a simulation study evaluating the performance of the existing and proposed methods. In Section \ref{Examples}, we illustrate these approaches on a real-life data set. We conclude with a discussion in Section \ref{Discussion}.

\section{Methods} \label{Methods}

\subsection{Standard Fixed Effect and Random Effects Models} \label{theory}

We first describe the standard fixed effect and random effects models for the meta-analysis of continuous outcomes.  For study $i$ where $i=1,\dots, k$, let $y_i$ denote the estimated effect measure and let $\sigma_i^2$ be its estimated variance.  For instance, the effect measure may be the difference of means or difference of medians of the outcome in two arms of a trial.  Note that although $\sigma_i^2$ is estimated from the data, it is typically assumed to be a known quantity in meta-analysis. The fixed effect model posits that a single true effect measure, $\theta$, underlies all studies.  Under standard assumptions, 
\begin{equation*}
y_i = \theta + \epsilon_i,
\end{equation*}
where $\epsilon_i \sim \mathcal{N}(0, \sigma_i^2)$ is random within-study variation. The pooled effect measure is then estimated by a weighted mean of $y_i$,
\begin{equation*}
\hat{\theta}_{FE} =  \frac{\sum \limits_{i=1}^k y_iw_{i} }{\sum \limits_{i=1}^k w_{i} }, \quad \widehat{Var}(\hat{\theta}_{FE})=\frac{1}{\sum \limits_{i=1}^k w_{i}}.
\end{equation*}
The weights for the estimates above are given by $w_i = 1/ \sigma_i^2$. 

The random effects model, on the other hand, presumes that different studies have different true effect measures.  The true effect measure of study $i$ is sampled from a normal distribution with mean $\theta$ and variance $\tau^2$.  That is, 
\begin{equation*}
y_i = \theta + \theta_i + \epsilon_i,
\end{equation*}
where $\theta_i \sim \mathcal{N}(0, \tau^2)$ is random between-study variation,  $\epsilon_i \sim \mathcal{N}(0, \sigma_i^2)$ is random within-study variation, and $\theta_i$ and $\epsilon_i$ are independent. The pooled effect measure is estimated by 
\begin{equation*}
\hat{\theta}_{RE} =  \frac{\sum \limits_{i=1}^k y_iw^*_{i} }{\sum \limits_{i=1}^k w^*_{i} }, \quad \widehat{Var}(\hat{\theta}_{RE})=\frac{1}{\sum \limits_{i=1}^k w^*_{i}}.
\end{equation*}   
The weights are given by $w^*_i=1/(\sigma_i^2 + \hat{\tau}^2)$ where $\hat{\tau}^2$ is an estimate of $\tau^2$. The method of moments estimator of DerSimonian and Laird \cite{dersimonian1986meta} is commonly used to estimate $\tau^2$, which is given by
\begin{equation*}
\hat{\tau}^2_{DL}= \max \left \{ 0,\frac{Q-(k-1)}{\sum \limits_{i=1}^k w_{i} - \sum \limits_{i=1}^k w_{i}^2 /  \sum \limits_{i=1}^k w_{i}}  \right \},
\end{equation*} 
where 
\begin{equation*}
Q =\sum \limits_{i=1}^k w_{i}(y_i-\hat{\theta}_{FE})^2. 
\end{equation*}
Under the typical assumption that the variances are known, the fixed effect and random effects estimators are minimum variance unbiased estimators \cite{rukhin2013}.

\subsection{Existing Methods: Transformation-Based Approaches} \label{scenarios}

For group $j$ ($j=1,2$) in study $i$, we use the following notation for summary measures that might be reported.  Let $n_{ij}$ be the number of subjects, $y_{ij}$ be the median of the outcome,  $q_{1,ij}$ and $q_{3,ij}$ be the first and third quartiles, respectively, and let $a_{ij}$ and $b_{ij}$ be the minimum and maximum values, respectively.  Furthermore, let $\bar x_{ij}$ be the sample mean and $s_{ij}$ be the sample standard deviation.  

We consider the following sets of summary measures that studies may report. In scenario 1 ($S_1$), a study reports the median, minimum and maximum values, and number of subjects.  In scenario 2 ($S_2$), a study reports the median, first and third quartiles, and number of subjects. Lastly, in scenario 3 ($S_3$), a study reports the median, minimum and maximum values, first and third quartiles, and number of subjects. 

We estimate the sample mean and standard deviation of the outcome in each group in each study. Then we estimate a fixed effect and random effects pooled difference of means and its 95\% confidence interval (CI) from the estimated differences of means and their variances \cite{hedges1998fixed}. For the random effects methods, we estimate $\tau^2$ using the method of DerSimonian and Laird \cite{dersimonian1986meta}. We call these approaches \emph{transformation-based approaches}.  In the following subsections, we describe the methods for estimating the sample mean and standard deviation under scenarios $S_1$, $S_2$, and $S_3$.

\subsubsection{Method of Wan et al}
Wan et al \cite{wan2014estimating} give an overview of methods for estimating the sample mean and standard deviation under scenarios $S_1$, $S_2$, and $S_3$. Wan et al \cite{wan2014estimating} recommend the following methods based on their simulation study. For $S_1$, Wan et al \cite{wan2014estimating} suggest using Hozo's method \cite{hozo2005estimating} to estimate the sample mean and recommend using their proposed method to estimate standard deviation. These estimates are given by:
\begin{align*}
\bar{x}_{ij} & \approx 
\begin{cases} 
\frac{a_{ij} + 2y_{ij} + b_{ij}}{4} &  \textrm{ if } n_{ij} \leq 25 \\ 
y_{ij} &   \textrm{ if } n_{ij} > 25
\end{cases} \\
s_{ij} & \approx \frac{b_{ij} - a_{ij}}{2\Phi^{-1}\left( \frac{n_{ij}-0.375}{n_{ij}+0.25} \right)},
\end{align*}
where $\Phi^{-1}$ denotes the inverse of the standard normal cumulative distribution function.  For $S_2$, Wan et al \cite{wan2014estimating} develop the following methods to estimate the sample mean and standard deviation:
\begin{align*}
\bar{x}_{ij}  & \approx \frac{q_{1,ij} + y_{ij} + q_{3,ij}}{3} \\
s_{ij} & \approx \frac{q_{3,ij} - q_{1,ij}}{2\Phi^{-1}\left( \frac{0.75n_{ij}-0.125}{n_{ij}+0.25} \right)}.
\end{align*}
Lastly, in $S_3$, Wan et al \cite{wan2014estimating} suggest using the method of Bland \cite{bland2014estimating} to estimate the sample mean and recommend using their method to estimate the standard deviation, which are given by:
\begin{align*}
\bar{x}_{ij}  & \approx \frac{a_{ij} + 2 q_{1,ij} + 2y_{ij} + 2q_{3,ij} + b_{ij}}{8} \\
s_{ij} & \approx  \frac{b_{ij} - a_{ij}}{4\Phi^{-1}\left( \frac{n_{ij}-0.375}{n_{ij}+0.25} \right)} + \frac{q_{3,ij} - q_{1,ij}}{4\Phi^{-1}\left( \frac{0.75n_{ij}-0.125}{n_{ij}+0.25} \right)}.
\end{align*}
A brief outline of the derivation of these methods is given below.  First, we consider estimating the sample mean in $S_1$ and $S_3$.  Hozo et al \cite{hozo2005estimating} and Bland \cite{bland2014estimating} place upper and lower bounds on the ordered observations with the reported sample quantiles. By summing the $n_{ij}$ inequalities and dividing by $n_{ij}$, one obtains bounds for the sample mean in $S_1$ and $S_3$. Then, one estimates the sample mean by the average of the upper and lower bounds. Note that these methods make no distributional assumptions for the outcome variable. 

To estimate the sample mean in $S_2$, Wan et al \cite{wan2014estimating} assume that the outcome distribution is normal. Wan et al \cite{wan2014estimating} show that the mean of the normal distribution can be expressed in terms of the expected value of the median and first and third quartiles. Then, Wan et al \cite{wan2014estimating} estimate the sample mean by replacing the expected value of the sample median and first and third quartiles with the observed sample values. 

Similarly, Wan et al \cite{wan2014estimating} estimate the sample standard deviation in $S_1$, $S_2$, and $S_3$ under the assumption that the outcome distribution is normal. Wan et al \cite{wan2014estimating} show that the standard deviation of the normal distribution can be expressed in terms of the expected value of sample quantiles. One can then estimate the sample standard deviation by replacing the expected value of the sample quantiles with the observed sample quantiles.

\subsubsection{Method of Luo et al}
We also consider the methods more recently recommended of Luo et al \cite{luo2016optimally}. Under the assumption that the outcome variable is normally distributed, the methods of Luo et al \cite{luo2016optimally} optimize the sample mean estimates recommended by Wan et al \cite{wan2014estimating}.  The formulas for estimating the sample mean from $S_1$, $S_2$, and $S_3$ are given below: 
\begin{equation*}
\bar x_{ij} \approx
\begin{cases}
 \left( \frac{4}{4+n_{ij}^{0.75}} \right) \frac{a_{ij} + b_{ij}}{2} + \left( \frac{n_{ij}^{0.75}}{4+n_{ij}^{0.75}} \right) y_{ij} &  \textrm{in $S_1$} \\ 
 \left( 0.7 + \frac{0.39}{n_{ij}} \right) \frac{q_{1,ij} + q_{3,ij}}{2} + \left( 0.3 - \frac{0.39}{n_{ij}} \right) y_{ij} & \textrm{in $S_2$}  \\ 
 \left( \frac{2.2}{2.2+n_{ij}^{0.75}}\right)\frac{a_{ij} + b_{ij}}{2} + \left(0.7 - \frac{0.72}{n^{0.55}} \right) \frac{q_{1,ij} + q_{3,ij}}{2} + \left(0.3 + \frac{0.72}{n^{0.55}} - \frac{2.2}{2.2+n^{0.75}} \right)y_{ij}& \textrm{in $S_3$}  \\ 
\end{cases}
\end{equation*}
To estimate the standard deviation, Luo et al \cite{luo2016optimally} recommend the methods previously described by Wan et al \cite{wan2014estimating}.

\subsection{Proposed Methods} \label{proposed methods}

We propose several methods to directly meta-analyze the difference of medians when the primary studies present summary statistics of $S_1$, $S_2$, or $S_3$. The target parameter in this case is the population difference of medians. These approaches, which we call \emph{median-based approaches}, fall into two categories.  In the first category, we consider simple approaches that do not apply the standard inverse variance method.  The second category of approaches estimates the variance of the difference of medians and then pools the studies using the standard inverse variance method described in Section \ref{theory}.  

\subsubsection{Median of the Difference of Medians} \label{median of medians}

McGrath et al \cite{mcgrath2018} proposed methods for pooling one-group studies that report the median of the outcome. These methods use the median of the study-specific medians as the pooled estimate of the median and invert the sign test to construct a CI around the pooled estimate. We consider extending the methods proposed by McGrath et al \cite{mcgrath2018} for pooling two-group studies as follows. We take a median of the study-specific differences of medians as the pooled estimate, and invert the sign test to construct the corresponding CI \cite{boldin1997}. Note that this method is nonparametric and the CI does not have coverage probability of exactly 95\% because of the discreteness of the sign test. Therefore, the coverage probability of the interval is slightly higher than 95\%, which depends on the number of studies included in the meta-analysis. We call this method the \textit{median of the difference of medians} (MDM) method.

If the study-specific difference of medians are drawn independently from a continuous distribution with median equal to the population difference of medians, the proposed method consistently estimates the target parameter. Moreover, the pooled estimator is a Hodges-Lehmann estimator in this case.

\subsubsection{Linear Quantile Mixed Models} \label{lqmm}
Quantile regression models estimate the conditional quantile of the outcome variable.  Linear quantile mixed models extend quantile regression models to include random effects \cite{geraci2014}. We apply linear quantile mixed models (LQMM) to meta-analyze the studies, where the the study-specific difference of medians are random intercepts in the model. The estimated fixed effect intercept and its 95\% CI are then used as the pooled estimate and its corresponding CI. 

The LQMM models were fit using the `lqmm' package in the R Statistical Software. Several numerical integration methods are available for model fitting, and different numerical integration methods lead to different distributional assumptions of the random intercepts. Because study-specific difference of medians are assumed to be normally distributed, Gauss-Hermite quadrature was chosen for numerical integration of the likelihood \cite{geraci2014}.

\subsubsection{Methods Based on Estimating the Variance of the Difference of Medians} \label{estvar}

We estimate the variance of the difference of medians of a given study using the asymptotic variance of the difference of medians.  For group $j$ in study $i$, let $f_{ij}$ be the probability density function of the underlying distribution, $m_{ij}$ be the study-population median, $y_{ij}$ be the sample median, and $n_{ij}$ be the number of subjects.  The asymptotic distribution of $y_{ij}$ is normal with mean $m_{ij}$ and variance $\frac{1}{4n_{ij}f_{ij}(m_{ij})^2}$.  It then follows that the asymptotic distribution of $y_i := y_{i1} - y_{i2} $ is
\begin{equation*}
y_i \sim \mathcal{N} \left( m_{i1} - m_{i2}, \frac{1}{4n_{i1}f_{i1}(m_{i1})^2} + \frac{1}{4n_{i2}f_{i2}(m_{i2})^2}  \right).
\end{equation*}  
We estimate the difference of medians, $m_{i1} - m_{i2}$, by replacing these quantities with the respective sample medians.  To estimate the variance of the difference of medians, we use
\begin{equation} \label{variance}
\widehat{Var}(y_i) = \frac{1}{4} \left( \frac{1}{n_{i1} \widehat{f_{i1}(m_{i1})}^2} + \frac{1}{n_{i2} \widehat{f_{i2}(m_{i2})}^2}\right),
\end{equation}
where $\widehat{ f_{i1} (m_{i1})}$ and $\widehat{f_{i2}(m_{i2})}$ are estimates of $f_{i1}(m_{i1})$ and $f_{i2}(m_{i2})$, respectively. In Section \ref{ls}, we propose a method to estimate these quantities for each study.  After estimating the difference of medians and its variance for each study, we estimate a pooled difference of medians using the standard meta-analytic methods presented in Section \ref{theory}.  As with the transformation-based methods, we use the method of DerSimonian and Laird \cite{dersimonian1986meta} to estimate heterogeneity.

For notational simplicity in the following section, we fix study $i$ and group $j$ and consequently drop the subscripts $i$ and $j$ from all relevant variables and functions.  For instance, we denote the density function of the underlying distribution evaluated at the study-population median of group $j$ of study $i$ as $f(m)$ instead of $f_{ij}(m_{ij})$.

\paragraph{Quantile Estimation \vspace{2mm} \\}  \label{ls}

Given an independent group in a primary study that reports summary statistics of $S_1$, $S_2$, or $S_3$, we propose the following method based on least squares to estimate the density function of the underlying distribution evaluated at the study-population median (i.e., $f(m)$).  First, we select a model parametrized by $\theta$ as the underlying distribution, which we denote as $P$.  Then, we fit the parameters of the distribution as follows.  We define $S_{P}(\theta)$ to be the sum of squares of the discrepancy between the theoretical quantiles of $P$ and the observed sample quantiles.  That is, in $S_1$, we use
\begin{equation*}
S_P(\theta) = (F^{-1}(1/n)-a)^2+ (F^{-1}(0.50)-y)^2+ (F^{-1} (1-1/n)-b)^2,
\end{equation*}
where $a$, $y$, and $b$ denote the observed sample minimum, median, and maximum, respectively. Moreover, $n$ denotes the sample size and $F(\cdot)$ denotes the cumulative distribution function of $P$. In $S_2$, $S_P(\theta) $ is expressed as
\begin{equation*}
S_P(\theta) = (F^{-1}(0.25)-q_1)^2+ (F^{-1}(0.50)-y)^2+ (F^{-1} (0.75)-q_3)^2,
\end{equation*}
where $q_1$ and $q_3$ denote the observed first quartile and third quartile, respectively. Lastly, in $S_3$,
\begin{align*}
S_P(\theta) = (F^{-1}(1/n)-a)^2 + (F^{-1}(0.25)-q_1)^2+ (F^{-1}(0.50)-y)^2 \\ + (F^{-1} (0.75)-q_3)^2+(F^{-1} (1-1/n)-b)^2.
\end{align*}
We then fit the parameters of $P$ by minimizing $S_P(\theta)$ with respect to $\theta$,
\begin{equation*}
\hat{\theta}=\operatorname*{arg\,min}_{\theta} S_P(\theta).
\end{equation*}

We consider that the underlying parametric family of distributions may be the normal, log-normal, Weibull, or gamma distribution.  These distributions were chosen to reflect a wide variety of possible outcome distributions. For each parametric family of distributions considered, we fit the parameters by minimizing $S_P(\theta)$.  The parametric family of distributions yielding the smallest value for $S_{P}(\hat{\theta})$ is chosen to be the underlying parametric family of distribution, thereby allowing us to obtain $\widehat{f(m)}$. We call this method \emph{quantile estimation} (QE).

To minimize $S_{P}(\theta)$ with respect to $\theta$, we use the limited-memory Broyden-Fletcher-Goldfarb-Shanno (BFGS) algorithm, denoted by L-BFGS-B, which is a quasi-Newton algorithm with box constraints \cite{Byrd1995}.  For the log-normal and normal distributions, the study-population median was constrained to be between the minimum and maximum values in $S_1$ and between the first and third quartiles in $S_2$ and $S_3$.  We allowed a wide range of values for $\sigma$ for the log-normal and normal distributions.  Additionally, for the Gamma and Weibull distribution, we considered a wide range of values for the parameters.  Table \ref{constraints} summarizes the parameter constraints in $S_1$, $S_2$, and $S_3$ for all distributions.

We imposed suitable constraints for the parameter values in the L-BFGS-B algorithm (e.g., not allowing negative variance). The specific values for the constraints were chosen based on the work of Kwon and Reis \cite{kwon2015simulation}. Specifically, in the Bayesian transformation-based method of Kwon and Reis, the authors used these values for the uniform prior bounds when fitting the normal, log-normal, and Weibull distributions in $S_1$, $S_2$, and $S_3$. For the gamma distribution, we set the constraints for the $\beta$ parameter based on the uniform prior bounds of Kwon and Reis for fitting the exponential distribution in $S_1$, $S_2$, and $S_3$, and we allowed a wide interval for the $\alpha$ parameter. 

We implemented the described L-BFGS-B algorithm using the optim function available in `stats' package in R Statistical Software. Convergence of the L-BFGS-B algorithm was defined as when the reduction of the objective function is within a factor of $10^{7}$ of machine tolerance, which corresponds to a tolerance of approximately $10^{-8}$. In all replications in the simulation study, the L-BFGS-B algorithm converged for at least one distribution. For each distribution, changes to the specific values of the constraints did not affect the solution provided the algorithm converged.

\paragraph{Best-Case Scenario for Quantile Estimation \vspace{2mm} \\}  \label{best-case}

As a best-case scenario for the quantile estimation method, we use the true value of $f(m)$ to estimate the variance of the difference of medians. That is, we estimate the variance of the difference of medians for each study using equation (\ref{variance}), where $\widehat{ f_{i1} (m_{i1})}$ and $\widehat{f_{i2}(m_{i2})}$ are replaced with the true values of $ f_{i1} (m_{i1})$ and $f_{i2}(m_{i2})$, respectively. We denote this method by QE-BC.

\subsection{Summary of Methods}

We provide a brief summary of all methods considered to meta-analyze two-group studies that report the median of the outcome of interest.  All methods fall into one of two categories: methods that meta-analyze the difference of means in order to estimate the population difference of means, and methods that meta-analyze the differences of medians in order to estimate the population difference of medians.

\begin{itemize}
\item[] Methods estimating the population difference of means:
\begin{itemize}
\item[] \textbf{Wan et al}: The mean and SD are estimated in each group in each study using the methods recommended by Wan et al \cite{wan2014estimating}.  Then, we meta-analyze the estimated difference of means using the inverse variance method described in Section \ref{theory}.
\item[] \textbf{Luo et al}: We use the method Luo et al \cite{luo2016optimally} to estimate the mean and SD in each group in each study.  We then pool the estimated  difference of means using the inverse variance method.
\end{itemize}
\item[] Methods estimating the population difference of medians:
\begin{itemize}
\item[] \textbf{Median of the Difference of Medians (MDM)}: We take the median of the study-specific difference of medians as the pooled estimate, and we invert the sign test to construct a CI around it. 
\item[] \textbf{Linear Quantile Mixed Models (LQMM)}: We fit a linear quantile mixed model using the study-specific difference of medians as random intercepts. The estimated fixed effect intercept and its 95\% CI is used as the pooled estimate and the corresponding CI.
\item[] \textbf{Quantile Estimation (QE)}: We estimate the variance of the difference of medians in each study using equation (\ref{variance}).  To estimate the density of the underlying distribution, we fit several distributions by minimizing the distance between the observed and theoretical quantiles and select the distribution with the best fit according to the proposed model selection procedure.  Then, we use the inverse variance method to meta-analyze the difference of medians.  
\item[] \textbf{Best-Case Scenario for Quantile Estimation (QE-BC)}:  We estimate the variance of the difference of medians in each study using (\ref{variance}), where the density of the underlying distribution is known. Then, we use the inverse variance method to pool the difference of medians.
\end{itemize}
\end{itemize}

In Appendix \ref{abc}, we describe a Bayesian method for density estimation and show the simulation results for this method. In Appendix \ref{modifications}, we describe several modifications to the proposed median-based methods and display their performance in the simulation study. These methods were excluded from the main paper because, in the simulation study, they did not considerably outperform the methods described in this section.

\section{Simulation Study} \label{Simulation}

\subsection{Data Generation}

We systematically compared the performance of the proposed median-based methods with the transformation-based methods via simulation study.  We simulated data for meta-analyzing two-group studies by varying the number of studies, number of subjects per study, underlying outcome distribution, and between-study heterogeneity.   

The number of studies was set to $k=10$ or $k=30$.  For a given study, the number of subjects was set to be equal across the two groups, which we denote as $n$.  The value for the number of subjects was drawn from a log-normal distribution with median either $50$ or $250$ and $\sigma=1$.  When the median number of subjects was set to $50$, we constrained the number of subjects to be between 10 and 500 by resampling until a value fell in the desired interval.  Similarly, when the median number of subjects was set to $250$, we constrained the number of subjects to be between 50 and 2,500.  

\begin{figure} [t] 
   \centering
    \caption{Probability density functions of the outcome distributions. Letting $f(x)$ denote the density of the mixture distribution, the mixture distribution can be described by $f(x) = \sum_{i=1}^{4} w_i \phi(x; \mu_i, \sigma_i^2)$, where we used $w_1 = 2/5$ and $w_2 = w_3 = w_4 = 1/6$ for the mixture probabilities (which were normalized to sum to 1), $\mu_1 = 36.5$, $\mu_2 = 40.5$, $\mu_3 = 44.5$, and $\mu_4 = 49.5$ for the means, and $\sigma_1 = 2.8$, $\sigma_2 = 3.6$, $\sigma_3 = 6$, and $\sigma_4 = 11$ for the standard deviations, respectively. \label{Distributions}}
 \includegraphics{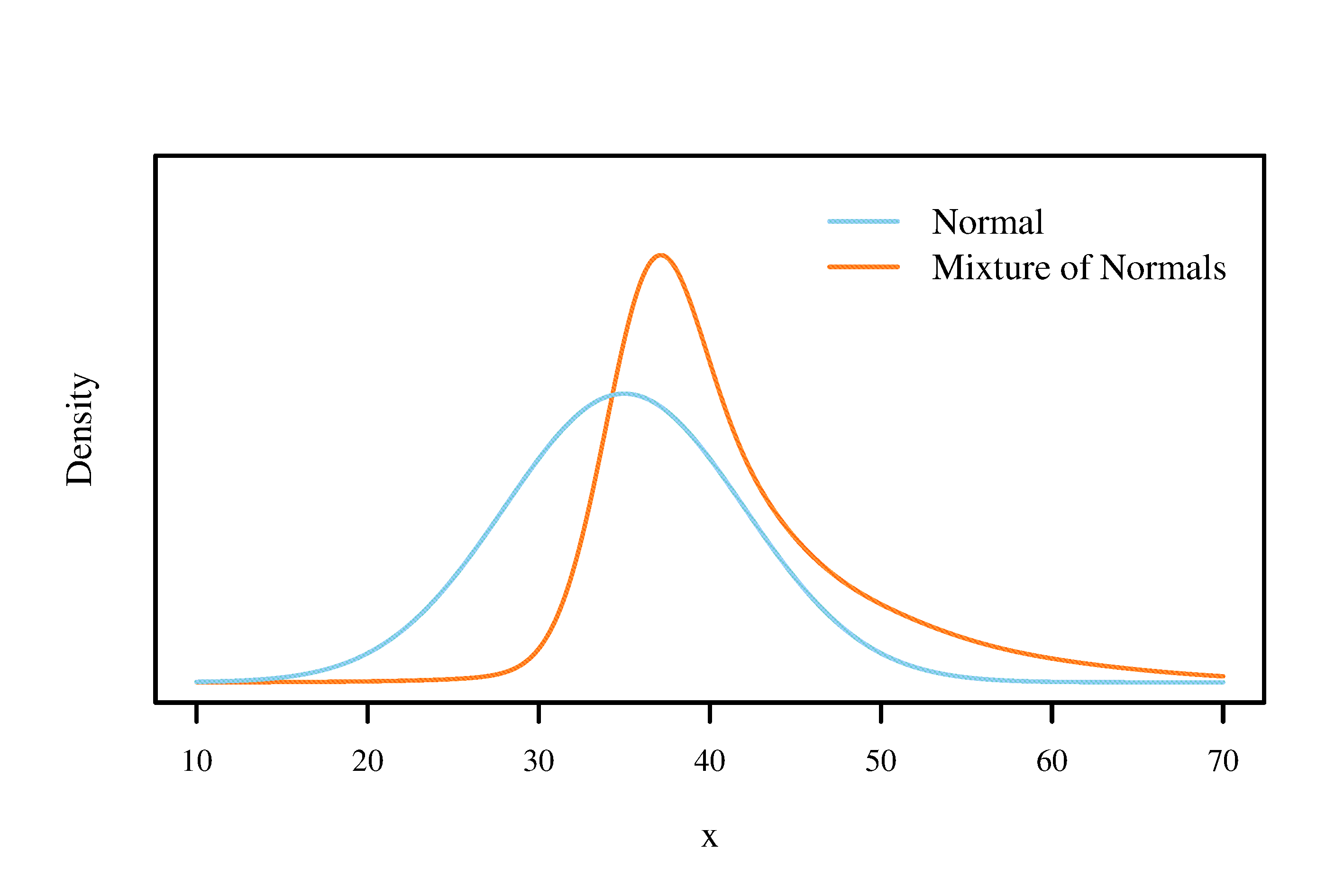}
 \end{figure}

We considered a total of three scenarios for the outcome distribution. In the first two scenarios, the outcomes for both groups were normally distributed and only differed by a location shift.  For each study, we sampled $n$ values from the normal distribution with $\mu=35$ and $\sigma=7$ two times independently, which we call the outcome values for groups 1 and 2.  An effect size, denoted by $c$, was then added to the outcome values in group 1 for all studies.  We considered a null and moderate effect size. The moderate effect size was chosen to obtain power of 0.60 in a two-sample z-test of the difference of means with $\alpha = 0.05$. For the third scenario, the group 1 outcome followed a moderately skewed mixture of normal distributions and the group 2 outcome followed a normal distribution with $\mu=35$ and $\sigma=7$ for all studies in the meta-analysis. The mixture of normal distributions had approximately a mean of 41, median of 39, and variance of 60. No additional effect size was added to the group 1 outcome values. Figure \ref{Distributions} displays the densities of the normal and mixture of normal distributions and gives the parameters values used for the mixture of normal distributions.

Lastly, we considered that the outcome either had no heterogeneity (i.e., a fixed effect model) or we added heterogeneity, denoted by $d$, which we generated from a normal distribution with mean $0$ and variance $\tau^2$.  When heterogeneity was included, we considered two levels of $\tau^2$ in order to obtain $I^2$ equal to 0.25 and 0.75---typically characterized as low and high levels of heterogeneity, respectively \cite{higgins2002quantifying}. For each study in a meta-analysis, we sampled a value of heterogeneity and added it to the group 1 outcome values. 

There were a total of $2 \times 2 \times 3 \times 3 = 36$ combinations of data generation parameters, summarized in Table \ref{parameters2}. For each of the 36 combinations, we simulated data for 1,000 meta-analyses.  

In our simulations, all primary studies in a given meta-analysis reported the same set of summary measures.  In the primary analysis, the primary studies reported summary measures of the outcome in the form of $S_1$, $S_2$,  and $S_3$, as described in Section \ref{scenarios}.

\subsection{Sensitivity Analysis: A Mix of Means and Medians} \label{sensitivity description}

When meta-analyzing real-life data, some primary studies may report the median of the outcome and others may report the sample mean.  We incorporated this scenario in our simulations as follows.  We considered that studies reported the median of the outcome in both groups if the Shapiro-Wilk normality test \cite{shapiro1965analysis} with significance level $\alpha=0.05$ is rejected in at least one group and reported the mean in both groups otherwise.  Additionally, we mimicked the scenario where authors do not consider the skewness of the outcome when choosing to report a mean or a median. Specifically, a random 25\%  of the primary studies reported the median and the remaining studies reported the mean of the outcome.  When studies reported the median of the outcome, the first and third quartiles and the number of subjects were also reported (i.e., $S_2$).  When the mean of the outcome was reported, the standard deviation and number of subjects were reported as well. For completeness, we also considered the scenario where all studies reported the sample mean, standard deviation, and number of subjects. This represents the best-case scenario for the transformation-based methods and the worst-case scenario for the median-based methods. 

We applied the transformation-based approaches and median-based approaches in these scenarios as follows.  For the transformation-based approaches, the sample mean and standard deviation were only estimated from studies that reported medians.  When studies reported the sample mean, standard deviation, and number of subjects, these values were used to calculate the difference of means and its variance.  

The quantile estimation method was applied as described in Section \ref{ls} when a study reported the median. When a study reported a mean, the method was applied in the following way.  We assumed that the outcome distribution was normal when studies reported the mean and fit the parameters of the normal distribution using the maximum likelihood estimates (i.e., $\hat \mu=\bar{x}$, $\hat \sigma^2 = \frac{n-1}{n}s^2$). Then, we estimated the variance of the difference of medians using equation (\ref{variance}).  

For the MDM and LQMM methods, the effect measure of the study was taken to be the difference of means for studies reporting means and the difference of medians for studies reporting medians.  Then, we took the median and its corresponding CI of the study-specific effect measures as the pooled median estimate for the MDM method. For LQMM method, the we applied the model as described in section \ref{lqmm}, where the random intercepts are the study-specific effect measures. Note that in this scenario, these approaches make the assumption that the sample means well-approximate the medians.

\subsection{Performance Measures}
We estimated the relative error (RE), variance, and coverage of the CIs of the pooled estimates. Letting $\hat{\theta}$ denote the estimate of the pooled effect measure and $\theta$ denote the true pooled effect measure, we define the relative error of $\hat{\theta}$ as 
\begin{equation*}
\textrm{RE}(\hat{\theta}) := \frac{\hat{\theta} - \theta }{\theta} \times 100.
\end{equation*}
We use the relative error of the pooled estimate as a performance measure rather than bias because, when the outcome follows a mixture of normals,  the value of the target parameter $\theta$ (i.e., the population difference of means for the transformation-based methods and the population difference in medians for the median-based methods) depends on the method used. The variance was calculated over the 1,000 meta-analyses for each of the 36 combinations of data generation parameters. 

Additionally, we calculated the bias of the estimate of heterogeneity. Letting $\hat{\tau}^2$ and $\tau^2$ denote the estimate and true value of heterogeneity, respectively, the bias was defined as $\textrm{Bias}(\hat{\tau}^2) := \hat{\tau}^2-\tau^2$.

We note that in the scenario where both group outcomes were normally distributed and the added effect size was equal to zero, the true pooled effect measure was equal to zero. In this scenario, we used bias instead of relative error as a performance measure.

\subsection{Results}

The results of the simulation study are given when the number of studies was 10, the median number of subjects per study was 50, the value of $I^2$ was $25\%$, and the added effect size in the normal distribution case was moderate. We found in preliminary analyses that these factors did not considerably affect the performance of the methods, and similar results hold for other fixed levels of these factors.  

In Appendix \ref{sensitivity}, we present the figures corresponding to the sensitivity analysis described in Section \ref{sensitivity description}. Moreover, we note that in some contexts data analysts prefer to conduct fixed effect analyses. Therefore, we present the results of the primary analysis where we apply the inverse variance methods in a fixed effect analysis in Appendix \ref{fixed effect}.

\subsubsection{Relative Error}

\begin{figure} [t] 
   \centering
    \caption{Relative error of the pooled estimates of the methods by the outcome distribution and summary measures reported. The methods were applied to the simulated data when the number of studies was 10, the median number of subjects per study was 50, the value of $I^2$ was $25\%$, and the added effect size in the normal distribution case was moderate. \label{re.primary}}
 \includegraphics[scale=0.9]{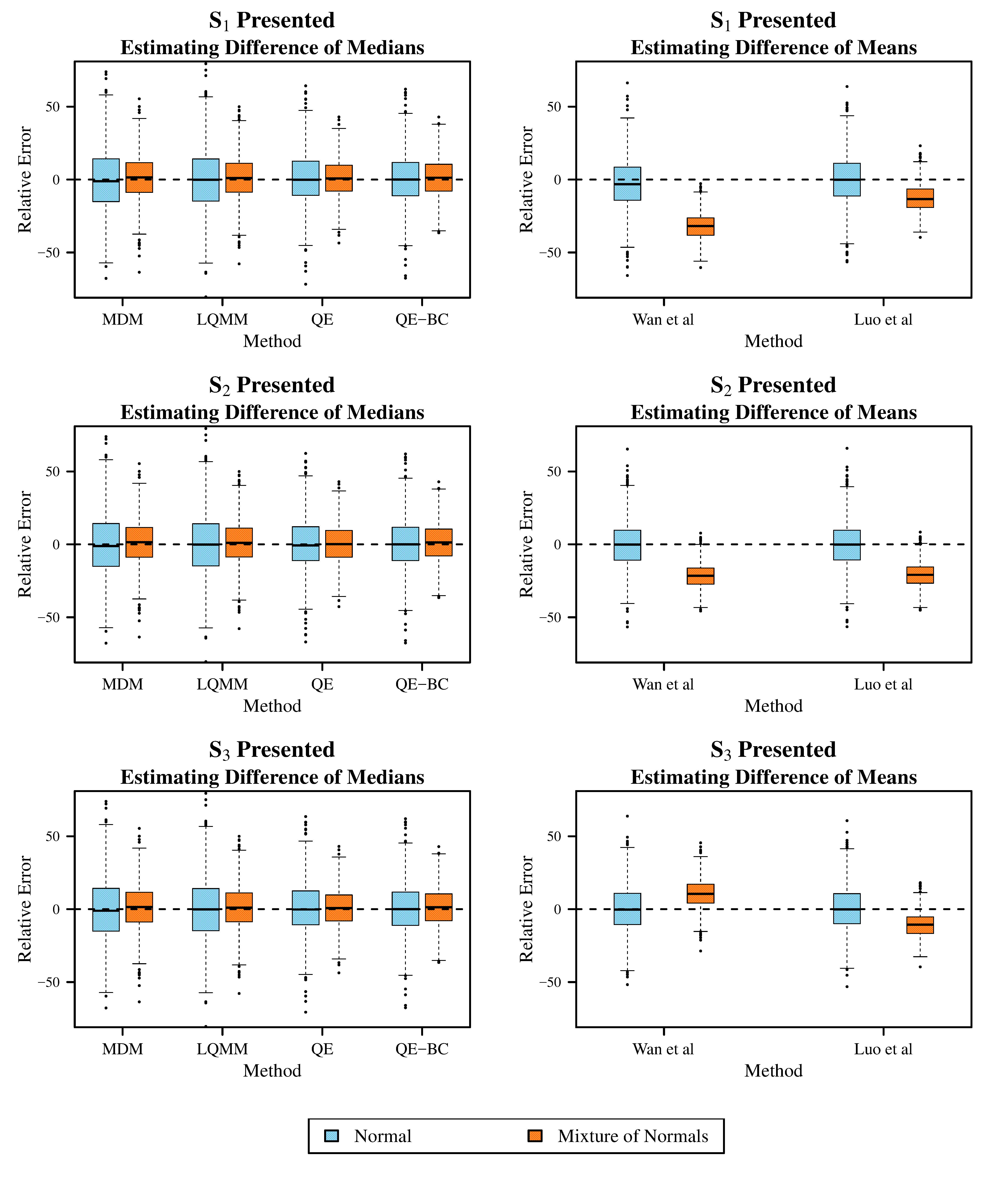}
 \end{figure}

Figure \ref{re.primary} displays the relative error of the pooled estimates by the outcome distribution and the summary measures reported in the primary analysis. The median-based methods have median relative error of nearly zero, regardless of the outcome distribution or summary measures given. The transformation-based methods also have median relative error of nearly zero for the normal distribution. However, the transformation-based methods display high relative error for the mixture of normals case. Although the method of Luo et al performs better than the method of Wan et al in $S_1$ when considering relative error, it does not offer a considerable improvement in $S_2$ or $S_3$ when the outcome distribution follows a mixture of normals.

Figure \ref{re.sensitivity} is the corresponding figure for the sensitivity analysis. When studies report the mean or median based on the skewness of the distribution, the relative error of all the approaches is similar to that of $S_2$. In the scenarios where (i) studies randomly report the mean or median of the outcome and (ii) all studies report the mean, all methods perform well when the outcome is normally distributed and the transformation-based methods outperform the median-based methods in the mixture case. 

 \subsubsection{Variance}

 Table \ref{table.variance} gives the variance of the methods by the outcome distribution and the summary measures reported. We first consider the primary analysis. The QE method has the smallest variance amongst the median-based methods, nearly equal to the variance when using the true density (i.e., QE-BC). Moreover, the variance of the QE method is not highly affected by the summary measures reported. The transformation-based methods have variance comparable to the QE method in the investigated scenarios. Although the methods of Luo et al and Wan et al have similar variance in most scenarios, the method of Luo et al outperforms that of Wan et al in regards to variance in $S_3$ when the outcome distribution follows a mixture of normals.

In all scenarios considered in the sensitivity analysis, the QE method has similar variance as the transformation-based methods as well as the QE-BC method. The other median-based methods (i.e., MDM, and LQMM) yielded higher values for the variance. Moreover, the variance of all approaches was not considerably affected by the summary measures reported in the sensitivity analysis.

 \subsubsection{Coverage}
\begin{figure} [t]
   \centering
    \caption{Coverage of the CIs of the pooled estimates of the methods by the outcome distribution and summary measures reported. The MDM method has coverage probability of 97.85\%, whereas all other methods have coverage probability of 95\%. The dotted line denotes coverage of 0.95. The methods were applied to the simulated data when the number of studies was 10, the median number of subjects per study was 50, the value of $I^2$ was $25\%$, and the added effect size in the normal distribution case was moderate.  \label{coverage.primary}}
 \includegraphics[scale=0.9]{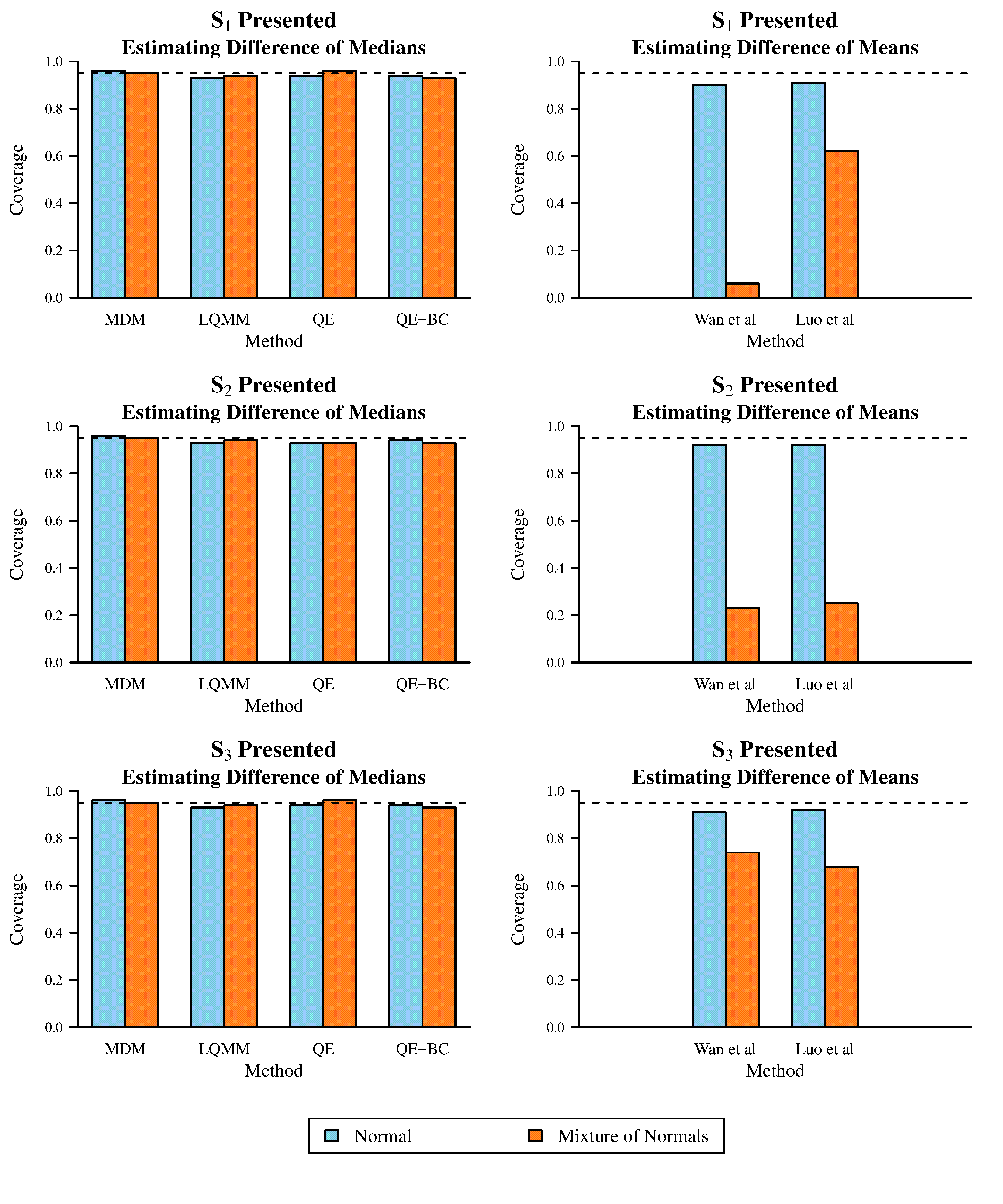}
 \end{figure}

Figure \ref{coverage.primary} gives the coverage of the CIs of the methods by the outcome distribution and the summary measures reported in the primary anlysis. Note that coverage probability of the CI for the MDM method depends on the number of studies, and coverage probability of this method is approximately 97.85\% for 10 studies. The median-based methods have nominal or near nominal coverage in all scenarios. For the transformation-based methods, their coverage is approximately 90\%-92\% in most scenarios where the outcome distribution was normal. However, the transformation-based method have poor coverage when the outcome distribution is a mixture of normals. The method of Luo et al only offers a considerable improvement to that of Wan et al when studies present $S_1$ and outcome distribution follows a mixture of normals.

Next, we consider the coverage of the methods in the sensitivity analysis, which is given in Figure \ref{cov.sensitivity}. In the scenario where studies report a mean or median based on the Shaprio-Wilk normality test, the coverage of all methods is similar to the $S_2$ scenario. When studies randomly report the mean or median, the transformation-based methods outperform the median-based methods in regards to coverage when the outcome is skewed. Specifically, when the outcome distribution is generated from the mixture of normals, the median-based methods have low coverage (i.e., less than 40\%) but the the transformation-based methods obtain coverage of approximately 88\%. The same conclusions hold for the scenario where all studies report sample means.

 \subsubsection{Bias for Estimating Heterogeneity}

\begin{figure} [t] 
   \centering
    \caption{Bias of the estimate of heterogeneity of the methods by the outcome distribution and summary measures reported. The methods were applied to the simulated data when the number of studies was 10, the median number of subjects per study was 50, the value of $I^2$ was $25\%$, and the added effect size in the normal distribution case was moderate. \label{ae.primary}}
 \includegraphics[scale=0.9]{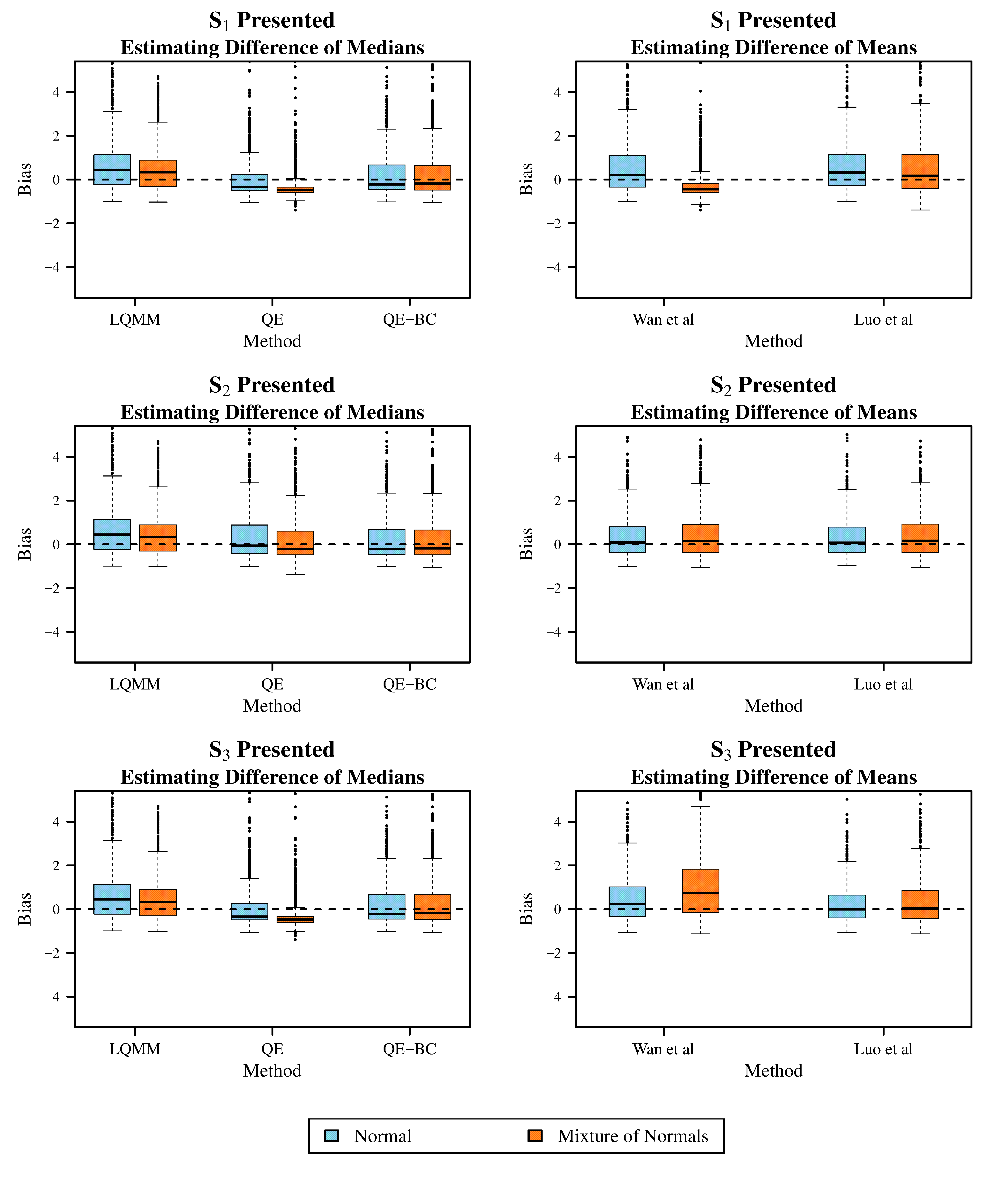}
 \end{figure}

Figure \ref{ae.primary} illustrates the bias for estimating heterogeneity by the outcome distribution and the summary measures reported. Note that the MDM method is not included because it is not a random effects approach. The QE-BC method performs better than the QE method for estimating heterogeneity when the outcome follows a mixture of normals in $S_1$ and $S_3$ and both methods perform comparably otherwise.  Similarly, for the transformation-based approaches, the method of Luo et al performs better than the method of Wan et al when the outcome follows a mixture of normals in $S_1$ and $S_3$. When comparing the transformation-based methods to the median-based methods, the method of Luo et al performs either better than or comparably to the LQMM and QE methods for estimating heterogeneity in all scenarios in Figure \ref{ae.primary}. 

Figure \ref{ae.sensitivity} displays the results for the bias for estimating heterogeneity in the sensitivity analysis. When studies report a mix of means and medians, the Luo et al and LQMM methods are preferable for estimating heterogeneity when the outcome is normal and the QE method performs best in the mixture case. When all studies report means, the method directly pooling the difference of means estimates heterogeneity better than the transformation-based methods, as expected.

\section{Example} \label{Examples}
To illustrate these approaches, we reanalyzed the meta-analysis conducted as part of the dissertation of Sohn \cite{hojoon2016improving}.  The goal of the meta-analysis is to evaluate the impact of novel diagnostic tests for reducing diagnostic delays for patients diagnosed for drug susceptible or drug resistant tuberculosis (TB). Diagnostic delay is defined in \cite{hojoon2016improving} as the length of time between an individual's first TB specific visit to a health care provider and the time of availability of diagnostic test results for clinical decision-making. The primary studies compared the delays experienced by two groups of patients, each receiving a different type of diagnostic test. One group received the Xpert MTB/RIF test, a cartridge-based nucleic acid amplification based test that can be performed at lower levels of the health system. Patients in the comparator group were diagnosed based on sputum smear microscopy, a century-old test widely used as the primary diagnostic test for TB. 

The relevant data set was extrapolated from nine studies, all of which reported the median, first quartile, and third quartile of the diagnostic delay along with the sample sizes in the Xpert and smear groups (i.e., $S_2$). Since diagnostic delay was measured in days, patients who received same-day test results were recorded as having a diagnostic delay of 0 days. In order to avoid complications for the quantile estimation method when fitting parameters of distributions with a strictly positive support, we added a value of 0.5 to all diagnostic delay summary data. Table \ref{description1} displays the relevant study-specific summary data.

In the original analysis, the author estimated the sample mean and standard deviation from the summary data using the methods recommended by Wan et al  \cite{wan2014estimating}. Then, they calculated the difference of means and its variance for each study, and they meta-analyzed the difference of means in a random effects analysis. 

In our analysis, we first apply transformation-based approaches to meta-analyze the data. We compare the methods recommended by Wan et al \cite{wan2014estimating} with the method of Luo et al \cite{luo2016optimally} for estimating the difference of means. We display the estimated study-specific difference of means and their 95\% CIs in Table \ref{effect.sizes.example}. We observe that the estimated difference of means and their 95\% CIs obtained by the two methods are nearly identical.

We next considered using the difference of medians as the effect measure. In this case, we apply the MDM, LQMM, and QE methods to pool the difference of medians, as described in Section \ref{proposed methods}. Table \ref{effect.sizes.example} displays study-specific difference of medians along with their estimated 95\% CIs using the QE method.  

We display the pooled estimates and their 95\% CIs of all considered methods in Table \ref{pooled1}. We also report the $I^2$ statistic \cite{higgins2002quantifying} and the $p$--value of the $\chi^2$ test of heterogeneity \cite{cochran1954} for the methods that pool studies with the inverse variance method (i.e., the Wan et al \cite{wan2014estimating}, Luo et al \cite{luo2016optimally}, and QE methods). Moreover, we report the estimate of heterogeneity for the random effect methods (i.e., the Wan et al \cite{wan2014estimating}, Luo et al \cite{luo2016optimally}, LQMM, and QE methods).

The estimated pooled difference of means of the two transformation-based methods are very similar and are centered at 2.1 days. The median-based methods estimate a pooled difference of medians of approximately 1 day. Moreover, the inverse variance methods all estimate significant heterogeneity. 

\section{Discussion} \label{Discussion}

We proposed several methods to meta-analyze two-group studies that report the median of the outcome along with the sample size and several measures of spread. The proposed methods directly pool the difference of medians, whereas the existing transformation-based methods \cite{wan2014estimating, hozo2005estimating, bland2014estimating, kwon2015simulation, kwon2016, wan2014estimating, luo2016optimally} are commonly used to pool estimates of the difference of means. 

Based on the results of the simulation study, we make the following suggestions to data analysts meta-analyzing two-group studies that report the median of the outcome. If all or nearly all studies report the median of the outcome, the median-based methods are expected to perform better than or comparably to the transformation-based methods. Similarly, if all or nearly all studies report the mean, the transformation-based methods should be used. In the case that some studies report the mean and others report the median, several additional considerations must be made. If the study-level outcome distribution is approximately normal, then all the methods considered in this paper are expected to perform well. If the study-level outcome distribution is highly skewed in some of the primary studies, then the transformation-based methods may be better suited. In practice, the skewness of the outcome distribution may be evaluated by Bowleys' coefficient of skewness \cite{kenney1962}--provided that the primary studies report the appropriate summary measures--or authors may make assumptions of skewness based on a priori domain knowledge of the outcome.  

Amongst the median-based methods, we recommend the quantile estimation method for most analyses. We found that this method performs better than the median of the difference of medians method and the linear quantile mixed model method in nearly all scenarios of the simulation study. Although we also implemented the Approximate Bayesian Computation method to improve density estimation by offering more well-established model selection and model averaging techniques (see Appendix \ref{abc}), we found that this approach performs nearly identically to the quantile estimation method in all scenarios in the simulation study. Therefore,  in practice, we recommend the use of the quantile estimation method over the Approximate Bayesian Computation method because of its simplicity and low computational cost. Moreover, we note that the error induced by density estimation does not considerably affect the performance of the quantile estimation method, which can be seen by the fact that it performs similar to the approach using the true underlying density.  The quantile estimation method offers several other advantages because it uses the inverse variance method to pool studies. First, under the standard assumption that the variances are known, the corresponding estimator is a minimum-variance unbiased estimator \cite{rukhin2013}. Second, it allows data analysts to perform standard follow-up analyses--such as heterogeneity modelling and cumulative meta-analyses--as well as graphical assessments--such as forest plots and funnel plots. 

In the one-sample case for meta-analyzing medians studied by McGrath et al \cite{mcgrath2018}, they generated a skewed outcome in their simulations using a log-normal distribution. When generating the outcome in this manner, they found that meta-analytic methods using inverse variance weighting yielded highly biased estimates due to the correlation between the effect measures and their variances, which is a well-documented problem in the literature (e.g., see  \cite{Shuster2010, Emerson1993}). Therefore, we chose to use a mixture of normal distributions to generate a skewed outcome in this work so that the effect measures are independent of their variances. When meta-analyzing data where this correlation is present, the median of the difference of medians and linear quantile mixed model method may be the preferred median-based methods because these methods weight studies equally. In future work, we intend to study more thoroughly the effect of correlation between effect sizes and their variance on the proposed median-based methods.

We also considered applying weighted versions of the median of the difference of medians and linear quantile mixed model methods, where studies were weighted proportionally to their sample size. As found in the one-sample case of McGrath et al \cite{mcgrath2018}, these weighted variations did not considerably improve the performance of the methods in the simulation study. Therefore, we excluded these approaches for parsimony. 

Several authors have recently developed methods to estimate the sample mean and standard deviation from the median, sample size, and several measures of spread \cite{wan2014estimating, hozo2005estimating, bland2014estimating, kwon2015simulation, kwon2016, luo2016optimally}. We decided to use the methods recommended by Wan et al \cite{wan2014estimating} and the subsequently optimized sample mean estimators of Luo et al \cite{luo2016optimally} in our work for because they have been shown to clearly outperform the methods of Hozo \cite{hozo2005estimating} and Bland \cite{bland2014estimating}. No known comparisons of the method of Luo et al \cite{luo2016optimally} with those of Kwon and Reis \cite{kwon2015simulation, kwon2016} have been performed. When comparing the methods recommended by Wan et al \cite{wan2014estimating} with those of Kwon and Reis \cite{kwon2015simulation, kwon2016}, the skewness of the outcome distribution and summary measures reported have a stronger influence over the performance of the estimators compared to the choice of method. Since the method recommended by Wan et al \cite{wan2014estimating} are more widely used in practice compared to those of Kwon and Reis \cite{kwon2015simulation, kwon2016}, we opted for the methods recommended by Wan et al \cite{wan2014estimating} in this work. 

Lastly, we revisit the data set originally collected and analyzed by Sohn \cite{hojoon2016improving}. Since all the primary studies reported the median of the outcome, the results of the simulation study suggest that the median-based methods are most suitable for meta-analyzing the data.  The preferred quantile estimation method estimates a pooled difference of medians of approximately 1 day [95\% CI: 0.2, 1.9] between the Xpert and smear groups.  This method also indicate that the primary studies included in the meta-analysis are highly heterogeneous ($I^2 = 97\%$), consistent with the conclusions obtained in the original analysis.

\bibliography{References}

\clearpage
\begin{table} [t] 
\caption {L-BFGS-B parameter constraints.} \label{constraints}
\begin{center}
\begin{tabular}{ c | c | c | c }
  \hline			
  Scenario & Distribution & $\theta_{1}$ & $\theta_{2}$ \\
  \hline
  $S_1$ & Normal & $\mu \in (a, b)$ & $\sigma \in (10^{-3},50)$  \\ 
  & Log-Normal & $\mu \in (\log(a), \log(b))$ & $\sigma \in (10^{-3}, 10)$  \\
  & Gamma & $\alpha \in (10^{-3}, 40)$ & $\beta \in (10^{-3}, 40)$ \\
  & Weibull & $\lambda \in (10^{-3}, 50)$ & $k \in (10^{-3}, 50)$ \\
  $S_2$ \& $S_3$ & Normal & $\mu \in (q_1, q_3)$ & $\sigma \in (10^{-3},50)$  \\
  & Log-Normal & $\mu \in (\log(q_1), \log(q_3))$ & $\sigma \in (10^{-3}, 10)$  \\
  & Gamma & $\alpha \in (10^{-3}, 40)$ & $\beta \in (10^{-3}, 40)$ \\
  & Weibull & $\lambda \in (10^{-3}, 50)$ & $k \in (10^{-3}, 50)$ \\
  \hline  
\end{tabular}
\end{center}
\end{table}

\clearpage
\begin{table} [t]
\caption {Data generation parameters. In the ``Outcome distributions" rows, $y_1$ denotes the group 1 outcome values and $y_2$ denotes the group 2 outcome values for each study in a given meta-analysis.} \label{parameters2}
\begin{center}
\begin{tabular}{l | l l l l}
  \hline
 Parameter &  Values & \\ \hline
No. of studies & 10 & 30 \\ 
Median sample size & $50$ & 250 \\ 
Outcome distributions &$y_1 \sim \mathcal{N}(35,7^2) + c + d$ &  $y_1$ generated from \textrm{mixture} $ + d$ \\ 
& $y_2 \sim \mathcal{N}(35,7^2)$ & $y_2 \sim \mathcal{N}(35,7^2)$ \\
 $c$  & 0 & Set to obtain $\textrm{power} = 0.6$ \\ 
 $d$ & 0 & Set to obtain $I^2=0.25$ & Set to obtain $I^2=0.75$  \\
   \hline
\end{tabular}
\end{center}
\end{table}

\clearpage
\begin{table}[t]
\caption {Variance of the pooled estimates of the methods by the outcome distribution and summary measures reported in the primary and sensitivity analyses. The methods were applied to the simulated data when the number of studies was 10, the median number of subjects per study was 50, the value of $I^2$ was $25\%$, and the added effect size in the normal distribution case was the moderate level.} \label{table.variance}
\begin{center}
\begin{tabular}{@{\extracolsep{6pt}}llllllll@{}}
  \hline
  & & \multicolumn{2}{c}{Transformation Methods} &  \multicolumn{4}{c}{Median-Based Methods}    \\ 
  \cline{3-4} \cline{5-8}
  Scenario & Distribution &  Wan et al & Luo et al &  MDM & LQMM & QE  & QE-BC \\ 
  \hline
 $S_1$ & Normal  & 0.31 & 0.29 & 0.46 & 0.48 & 0.32 & 0.31 \\ 
 & Mixture  & 0.28 & 0.32 & 0.42 & 0.40 & 0.29 & 0.29 \\ 
   $S_2$ & Normal & 0.26 & 0.26 & 0.46 & 0.48 & 0.32 & 0.31 \\ 
  & Mixture  & 0.27 & 0.27 & 0.42 & 0.40 & 0.30 & 0.29 \\
  $S_3$ & Normal  & 0.27 & 0.24 & 0.46 & 0.48 & 0.32 & 0.31 \\ 
  & Mixture  & 0.37 & 0.27 & 0.42 & 0.40 & 0.29 & 0.29 \\ 
  Mix (S-W Test) & Normal  & 0.22 & 0.22 & 0.33 & 0.31 & 0.23 & 0.23 \\ 
  & Mixture  & 0.26 & 0.26 & 0.44 & 0.42 & 0.30 & 0.29 \\ 
   Mix (Random) & Normal  & 0.23 & 0.22 & 0.33 & 0.34 & 0.23 & 0.24 \\ 
  & Mixture & 0.26 & 0.26 & 0.42 & 0.41 & 0.27 & 0.27 \\ 
   Means & Normal  & 0.21 & 0.21 & 0.32 & 0.30 & 0.21 & 0.21 \\ 
  & Mixture  & 0.25 & 0.25 & 0.39 & 0.36 & 0.25 & 0.25 \\ 
  \hline
\end{tabular}
\end{center}
\end{table}

\clearpage
\begin{table} [t]
\caption {Study-specific summary data.  The format of the diagnostic delay summary data is (first quartile, median, third quartile), denoted by $(q_1, y, q_3)$ as in Section \ref{ls}. The unit of diagnostic delay is days.} \label{description1}
\begin{center}
\begin{tabular}{@{\extracolsep{6pt}}lllll@{}}
  \hline
  & \multicolumn{2}{c}{Smear} &  \multicolumn{2}{c}{Xpert}   \\ 
  \cline{2-3} \cline{4-5} 
Study & $n$ & $(q_1, y, q_3)$ & $n$ & $(q_1, y, q_3)$ \\ 
  \hline
Boehme et al 2011 & 3659 & (2.50, 2.50, 3.50) & 1429 & (0.50, 1.50, 3.50) \\ 
  Yoon et al 2012 & 190 & (0.50, 1.50, 26.50) & 246 & (0.50, 0.50, 26.50) \\ 
  Kwak et al 2013 & 681 & (7.50, 12.50, 19.75) & 681 & (3.50, 6.50, 19.75) \\ 
  Sohn et al 2014 & 11 & (1.50, 1.58, 2.60) & 11 & (0.50, 1.54, 2.60) \\ 
  Mupfumi et al 2014 & 210 & (1.50, 6.50, 25.50) & 214 & (1.50, 2.50, 25.50) \\ 
  Lippincott et al 2014 & 207 & (0.50, 1.83, 2.50) & 207 & (0.50, 1.25, 2.50) \\ 
  Durovni et al 2014 & 831 & (5.40, 8.00, 10.50) & 1385 & (3.90, 7.80, 10.50) \\ 
  Cohen et al 2014 & 90 & (2.60, 3.80, 5.70) & 156 & (5.80, 6.80, 5.70) \\ 
  Chaisson et al 2014 & 142 & (1.50, 2.50, 4.50) & 142 & (0.50, 1.50, 4.50) \\ 
   \hline
\end{tabular}
\end{center}
\end{table}

\clearpage
\begin{table} [t]
\caption {Study-specific (estimated) effect measures and their 95\% CIs. For the transformation-based methods, the effect measure is the estimated difference of means. For the QE method, the effect measure is the difference of medians.} \label{effect.sizes.example}
\begin{center}
\begin{tabular}{@{\extracolsep{6pt}}llll@{}}
\hline
  & \multicolumn{2}{c}{Transformation-based methods} &  \multicolumn{1}{c}{Median-based methods}   \\ 
  \cline{2-3} \cline{4-4} 
\multicolumn{1}{c}{Study} & \multicolumn{1}{c}{Wan et al}  & \multicolumn{1}{c}{Luo et al}  & \multicolumn{1}{c}{QE} \\ 
  \hline
Boehme et al 2011 & 1.33 [1.25, 1.41] & 1.35 [1.27, 1.43] & 1.00 [0.90, 1.10]  \\ 
  Yoon et al 2012 & 8.67 [6.24, 11.09] & 9.07 [6.64, 11.49] & 1.00 [-0.03, 2.03]  \\ 
  Kwak et al 2013 & 7.42 [6.70, 8.14] & 7.49 [6.77, 8.21] & 6.00 [5.11, 6.89]  \\ 
  Sohn et al 2014 & -0.25 [-2.28, 1.78] & -0.28 [-2.31, 1.75] & 0.04 [-1.84, 1.93] \\ 
  Mupfumi et al 2014 & 5.33 [2.63, 8.04] & 5.40 [2.70, 8.11] & 4.00 [1.51, 6.49]  \\ 
  Lippincott et al 2014 & 0.53 [0.30, 0.76] & 0.53 [0.30, 0.75] & 0.59 [0.30, 0.87] \\ 
  Durovni et al 2014 & 0.90 [0.56, 1.24] & 0.94 [0.60, 1.27] & 0.20 [-0.22, 0.62]  \\ 
  Cohen et al 2014 & -3.03 [-3.62, -2.45] & -3.03 [-3.62, -2.45] & -3.00 [-3.71, -2.29]  \\
  Chaisson et al 2014 & 1.33 [0.89, 1.78] & 1.35 [0.91, 1.80] & 1.00 [0.46, 1.54]  \\ 
   \hline
\end{tabular}
\end{center}
\end{table}

\clearpage
\begin{table} [t]
\caption {Pooled estimates of all methods. The pooled estimates and the 95\% CIs are given in the ``Pooled Estimate" column. We also display the estimated heterogeneity of the random effects methods, denoted by $\hat{\tau}^2$, the $I^2$ statistic, and the $p$--value of the $\chi^2$ test of heterogeneity.} \label{pooled1}
\begin{center}
\begin{tabular}{lllll}
  \hline
Method & Pooled Estimate & $\hat{\tau}^2$ & $I^2$(\%) & $p$--value  \\ 
  \hline
Wan et al & 2.08  [1.02, 3.14] & 2.24 &  98.63 & $p < 10^{-5}$  \\ 
  Luo et al & 2.14  [1.07, 3.21] & 2.29 & 98.66 & $p < 10^{-5}$   \\ 
  MDM & 1.00  [0.16, 4.22] &  &  &  \\ 
  LQMM & 1.00 [-0.45 , 2.45] & 1.00  &  &  \\ 
  QE & 1.05  [0.18, 1.91] & 1.49 & 96.99 & $p < 10^{-5}$ \\ 
   \hline
\end{tabular}
\end{center}
\end{table}

\clearpage

\clearpage

  \clearpage
 
  \clearpage

\appendix
\counterwithin{figure}{section}
\counterwithin{table}{section}

\clearpage
\section{Density Estimation via Approximate Bayesian Computation} \label{abc}

In this section, we propose a Bayesian method for density estimation, which offers more well-established model selection and model averaging techniques. Using the same notation as in Section \ref{ls}, we apply approximate Bayesian computation (ABC) to estimate $f(m)$ from $S_1$, $S_2$, and $S_3$.  In standard Bayesian analyses, the posterior distribution is expressed as
\begin{equation*}
p(\theta | y) \propto p(y | \theta) p(\theta),
\end{equation*}
where $p(y | \theta)$ denotes the likelihood and $p(\theta)$ denotes the prior distribution.  The main idea behind ABC is that a random sample drawn from a correctly specified likelihood should be close to the observed data. In ABC, one samples parameter values from $p(\theta)$, denoted by $\theta^*$, and then draws a sample from $p(y | \theta^*)$. We consider $\theta^*$ to be a draw from the approximate posterior distribution if the distance between the observed data and the generated pseudodata is sufficiently small. Specifically, we implement the following algorithm, adapted from the methods of Kwon and Reis \cite{kwon2015simulation, kwon2016}.

We begin with specifying $k$ candidate parametric families of distributions for the outcome variable, which we denote as $P_1, \dots, P_k$.  Let model $P_i$ be parametrized by $\theta_i$ for $i=1, \dots, k$.  We select a parametric family of distributions from a multinomial distribution with probabilities $p=(p_1, \dots, p_k)$ where initially $p_i=\frac{1}{k}$ for $i=1, \dots, k$.  Suppose model $P_i$ chosen for some fixed $i$.  We sample  $\theta^*$ from the specified prior distributions corresponding to model $P_i$.  Then we simulate a data set, $y^*$, with $n$ observations from the likelihood and compute summary statistics in the same form of the original study (i.e.,  $S_1$, $S_2$, or $S_3$).  If the distance between the summary statistics of $y^*$ and the study-specific summary statistics is sufficiently small, we retain $\theta^*$.  After every 1,000 iterations, we update the vector $p$ so that $p_i$ is the percentage of accepted parameter values for the model $P_i$.  After $N$ repetitions for some large $N$, the distribution of the accepted $\theta^*$ approximate the posterior distribution of model $P_i$.  We fit $\theta_i$ using the means of their respective posterior distributions.  

We consider applying ABC with (i) single distribution selection and (ii) Bayesian model averaging to estimate $f(m)$. As used by Kwon and Reis \cite{kwon2016}, we denote the ABC method with single distribution selection as $\textrm{ABC}_{\textrm{SDS}}$ and denote the ABC method with Bayesian model averaging as $\textrm{ABC}_{\textrm{BMA}}$.

For single distribution selection, we use Bayes' factor for model selection. That is, we select the distribution with the highest posterior model probability.  Kwon and Reis \cite{kwon2015simulation} show that the posterior model probability of model $P_i$ is well-approximated by the percentage of accepted parameter values for that model, $p_i$.   After selecting the parametric family of distributions with the highest posterior model probability, we estimate $f(m)$ using the density function and distribution median of the fitted distribution. 

For Bayesian model averaging, let $\widehat{ f(m)}^{(i)}$ denote the probability density function of the fitted distribution evaluated at its distribution median under model $i$ by applying the ABC algorithm.  Then
\begin{equation*}
\widehat{ f ( m)} = \sum_{i=1}^k p_i \widehat{ f(m)}^{(i)},
\end{equation*}
where $p_i$ is the estimated posterior model probability of model $P_i$.

Consistent with the methods of Kwon and Reis \cite{kwon2015simulation, kwon2016}, the distance between summary statistics of the simulated and observed data is measured in the $\ell_2$ norm.  Moreover, in standard practice, one often uses a small acceptance rate of sampled parameter values rather than a minimum threshold to select candidate parameter values \cite{beaumont2002}.  Specifically, we fix an acceptance rate of sampled parameters at 0.1\% and set the number of iterations to 20,000 for a given parametric family of distributions, as recommended in this context by Kwon and Reis \cite{kwon2015simulation, kwon2016}. That is, the 20 sample parameter values yielding summary statistics closest to the observed data were used to model the posterior distributions.  As with the least squares method, we consider the following candidate distributions: the normal, log-normal, Weibull, and gamma distributions.  Table \ref{priors} displays the prior distributions under $S_1$, $S_2$, and $S_3$. As with the box constraints used in the L-BFGS-B algorithm, these prior distribution were selected based on the ABC algorithms of Kwon and Reis \cite{kwon2015simulation, kwon2016}.

Because of the high computation cost of the ABC methods, we apply these methods to the first 500 (out of the 1,000) simulated data sets for each combination of data generation parameters. In Table \ref{table.abc}, we display the relative error of the pooled estimates, variance of the pooled estimates, coverage of the 95\% CIs of the pooled estimates, and bias of the heterogeneity estimates when the ABC methods were applied to the data generated in the primary analysis of the simulation study. In all scenarios, the ABC methods perform comparably to the QE method and its best-case scenario (i.e., the QE-BC method).

\clearpage
\begin{table}[t]
\caption {Prior distributions for ABC methods} \label{priors}
\begin{center}
\begin{tabular}{c | c | c | c | c | c}
  \hline			
  Scenario & Likelihood Distribution  & $\theta_{1}$ & $p(\theta_1)$ & $\theta_{2}$ & $p(\theta_2)$\\
  \hline
  $S_1$ & Normal & $\mu$ & $\mathcal{U}(a, b)$ & $\sigma$ & $\mathcal{U} (0,50)$  \\
  & Log-Normal & $\mu$ & $\mathcal{U}(\log(a), \log(b))$ & $\sigma$ & $\mathcal{U}(0, 10)$  \\
  & Gamma & $\alpha$ & $\mathcal{U}(0, 40)$ & $\beta$ & $\mathcal{U} (0, 40)$ \\
  & Weibull & $\lambda$ & $\mathcal{U}(0, 50)$ & $k$ & $\mathcal{U}(0, 50)$ \\
  $S_2$ \& $S_3$ & Normal & $\mu$ & $\mathcal{U}(q_1, q_3)$ & $\sigma$ & $\mathcal{U} (0,50)$  \\
  & Log-Normal & $\mu$ & $\mathcal{U}(\log(q_1), \log(q_3))$ & $\sigma$ & $\mathcal{U}(0, 10)$  \\
  & Gamma & $\alpha$ & $\mathcal{U} (0, 40)$ & $\beta$ & $\mathcal{U}(0, 40)$ \\
  & Weibull & $\lambda$ & $\mathcal{U} (0, 50)$ & $k$ & $\mathcal{U} (0, 50)$ \\
  \hline  
\end{tabular}
\end{center}
\end{table}

\clearpage
\begin{landscape}

\begin{table}[t]
\caption {Performance of the ABC methods compared to the QE and QE-BC methods. Performance is measured by relative error of the pooled estimate, variance of the pooled estimate, coverage of the 95\% CIs of the pooled estimate, and bias of the heterogeneity estimate. In the ``Relative Error" and ``Bias for Estimating Heterogeneity" columns, the data are presented in the format: median (first quartile, third quartile). The methods were applied to the simulated data when the number of studies was 10, the median number of subjects per study was 50, the value of $I^2$ was $25\%$, and the added effect size in the normal distribution case was moderate.  All methods were applied to the same 500 simulated data sets for each combination of data generation parameters.} \label{table.abc}
\begin{center}
\footnotesize
\begin{tabular}{@{\extracolsep{6pt}}llllllllll@{}}
  \hline
  & & \multicolumn{2}{c}{Relative Error} &  \multicolumn{2}{c}{Variance}  & \multicolumn{2}{c}{Coverage} & \multicolumn{2}{c}{Bias for Estimating Heterogeneity}   \\ 
  \cline{3-4} \cline{5-6} \cline{7-8} \cline{9-10}
  Scenario & Method &  Normal & Mixture &  Normal & Mixture & Normal  & Mixture & Normal  & Mixture \\ 
  \hline
$S_1$ & QE & -0.77 (-11.80, 13.13) & 0.92 (-7.72, 9.64) & 0.32 & 0.29 & 0.94 & 0.96 & -0.35 (-0.49, 0.18) & -0.49 (-0.61, -0.35) \\ 
  & $\textrm{ABC}_{\textrm{SDS}}$ & -1.14 (-11.79, 12.42) & 0.55 (-7.64, 9.90) & 0.33 & 0.29 & 0.92 & 0.95 & -0.26 (-0.45, 0.61) & -0.45 (-0.58, -0.22) \\
  & $\textrm{ABC}_{\textrm{BMA}}$ & -1.09 (-11.71, 12.34) & 0.58 (-7.91, 9.92) & 0.34 & 0.29 & 0.92 & 0.95 & -0.26 (-0.45, 0.60) & -0.46 (-0.58, -0.22) \\ 
  & QE-BC & -0.90 (-11.71, 11.80) & 1.47 (-7.61, 10.68) & 0.32 & 0.28 & 0.93 & 0.93 & -0.20 (-0.44, 0.65) & -0.16 (-0.50, 0.63) \\ 
  $S_2$ & QE & -1.85 (-12.07, 12.05) & 0.44 (-8.93, 9.26) & 0.34 & 0.29 & 0.93 & 0.92 & -0.05 (-0.41, 0.90) & -0.17 (-0.48, 0.58) \\
  & $\textrm{ABC}_{\textrm{SDS}}$ & -1.54 (-12.30, 11.09) & 0.27 (-9.08, 9.45) & 0.34 & 0.29 & 0.93 & 0.92 & -0.11 (-0.41, 0.84) & -0.24 (-0.50, 0.53) \\
  & $\textrm{ABC}_{\textrm{BMA}}$ & -1.62 (-12.08, 10.99) & 0.24 (-9.10, 9.42) & 0.34 & 0.29 & 0.93 & 0.92 & -0.10 (-0.41, 0.85) & -0.24 (-0.50, 0.56) \\
  & QE-BC & -0.90 (-11.71, 11.80) & 1.47 (-7.61, 10.68) & 0.32 & 0.28 & 0.93 & 0.93 & -0.20 (-0.44, 0.65) & -0.16 (-0.50, 0.63) \\ 
  $S_3$ & QE & -0.84 (-11.51, 12.77) & 0.78 (-7.78, 9.80) & 0.32 & 0.29 & 0.94 & 0.96 & -0.34 (-0.49, 0.27) & -0.48 (-0.61, -0.34) \\ 
  & $\textrm{ABC}_{\textrm{SDS}}$ & -1.26 (-11.77, 12.35) & 0.20 (-7.84, 9.97) & 0.34 & 0.29 & 0.92 & 0.95 & -0.18 (-0.42, 0.76) & -0.42 (-0.56, 0.02) \\ 
  & $\textrm{ABC}_{\textrm{BMA}}$ & -1.36 (-11.72, 12.30) & 0.15 (-7.61, 9.93) & 0.34 & 0.29 & 0.92 & 0.95 & -0.18 (-0.41, 0.78) & -0.42 (-0.56, 0.01) \\ 
  & QE-BC & -0.90 (-11.71, 11.80) & 1.47 (-7.61, 10.68) & 0.32 & 0.28 & 0.93 & 0.93 & -0.20 (-0.44, 0.65) & -0.16 (-0.50, 0.63) \\
   \hline
\end{tabular}
\end{center}
\end{table}
\end{landscape}

 \clearpage
 \section{Modifications to the Proposed Methods} \label{modifications}
In the following subsections, we discuss several modifications to the proposed median-based methods and evaluate the performance of the methods under these modifications. 

\subsection{Median of the Difference of Medians}

In the main paper, we invert the sign test to construct a CI around the pooled estimate of the MDM method. When the number of studies is sufficiently large, one can use the normality approximation of the binomial distribution to construct an approximate 95\% CI around the pooled estimate. Specifically, let $k$ denote the number of studies and $z_{0.025}$ denote the 0.975 quantile of the standard normal distribution.  The $\frac{1}{2}-\min \left( \frac{1}{2}, \frac{z_{0.025}}{2\sqrt{k}} \right)$ and $\frac{1}{2}+\min \left( \frac{1}{2}, \frac{z_{0.025}}{2\sqrt{k}} \right)$ quantiles of the study-specific difference of medians are the lower and upper limits, respectively, of the approximate 95\% CI of the MDM method \cite{conover1980practical}. We denote the MDM method with the normality approximation as MDM-N. 

We applied the MDM-N method to the data generated in our primary simulation study, and we compare its coverage and CI lengths to the MDM method. Theoretically, the coverage probability of the MDM method is approximately 97.85\% for 10 studies and 95.72\% for 30 studies. In Table \ref{table.b1}, we present the coverage and mean length of the CIs of these methods. In all scenarios considered, the MDM-N had nearly equal coverage and smaller mean length of CIs when compared to to the MDM method.

\subsection{Modifications for Density Estimation Methods}
In some contexts, authors may assume that the underlying outcome distribution of the two groups in a primary study only differ by a location shift. In this case, one can gain greater precision estimating the variance of the difference of medians by averaging the density estimates over the two groups. This method is described in the following paragraph.  

Using the notation of Section \ref{estvar}, we assume that $f_{i1}(m_{i1})=f_{i2}(m_{i2})$.  Under the location shift assumption, a new estimate of the variance of the difference of medians is given by
\begin{equation}  \label{variance2}
\widehat{Var}(y_i) = \frac{1}{4\widehat{f(m)}_i^2} \left( \frac{1}{n_{i1}} + \frac{1}{n_{i2}}\right)
\end{equation}
where $\widehat{f(m)}_i$ is given by
\begin{equation*}
\widehat{f(m)}_i = \frac{n_{i1}\widehat{f_{i1}(m_{i1})} + n_2 \widehat{f_{i2}(m_{i2})}}{n_{i1}+n_{i2}}.
\end{equation*}

However, we found in our simulations the accuracy of the pooled estimates of the density estimation methods using (\ref{variance2}) did not considerably improve the accuracy of the methods (data not shown).

\clearpage

\begin{table}[t]
\caption {Effect of normality approximation on the coverage and mean length of CIs for the MDM method. The methods were applied to the simulated data when the median number of subjects per study was 50, the value of $I^2$ was $25\%$, and the added effect size in the normal distribution case was moderate.  All primary studies reported the median of the outcome and the sample size in both groups. } \label{table.b1}
\begin{center}
\begin{tabular}{llllll}
  \hline
  & & \multicolumn{2}{c}{Coverage}  & \multicolumn{2}{c}{Mean Length of CIs}  \\ 
  \cline{3-4} \cline{5-6}
 N. Studies & Method & Normal & Mixture  & Normal & Mixture \\ 
  \hline
10 & MDM  & 0.95 & 0.94 & 3.37 & 3.24 \\ 
  & MDM-N  & 0.95 & 0.94 & 2.82 & 2.66 \\ 
  30 & MDM  & 0.95 & 0.96 & 1.62 & 1.51 \\
  & MDM-N & 0.95 & 0.95 & 1.58 & 1.47 \\ 
   \hline
\end{tabular}
\end{center}
\end{table}

\clearpage
\section{Sensitivity Analysis: A Mix of Means and Medians} \label{sensitivity}

\begin{figure} [H] 
   \centering
    \caption{Relative error of the pooled estimates of the methods by the outcome distribution and summary measures reported in the sensitivity analysis. The methods were applied to the simulated data when the number of studies was 10, the median number of subjects per study was 50, the value of $I^2$ was $25\%$, and the added effect size in the normal distribution case was moderate. \label{re.sensitivity}}
 \includegraphics[scale=0.9]{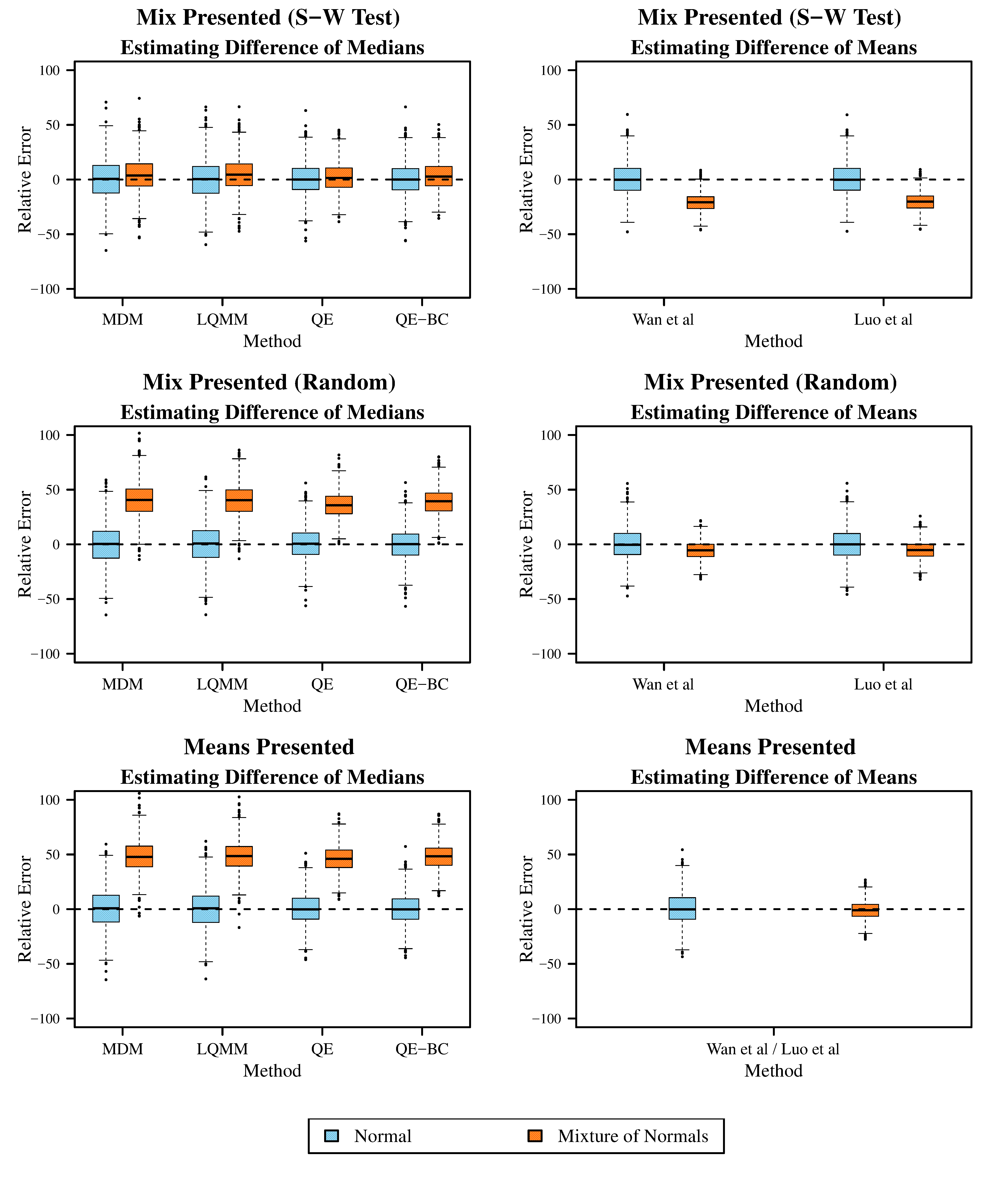}
 \end{figure}

\begin{figure} [H] 
   \centering
    \caption{Coverage of the CIs of the pooled estimates of the methods by the outcome distribution and summary measures reported in the sensitivity analysis. The MDM method has coverage probability of 97.85\%, whereas all other methods have coverage probability of 95\%. The dotted line denotes coverage of 0.95. The methods were applied to the simulated data when the number of studies was 10, the median number of subjects per study was 50, the value of $I^2$ was $25\%$, and the added effect size in the normal distribution case was moderate. \label{cov.sensitivity}}
 \includegraphics[scale=0.9]{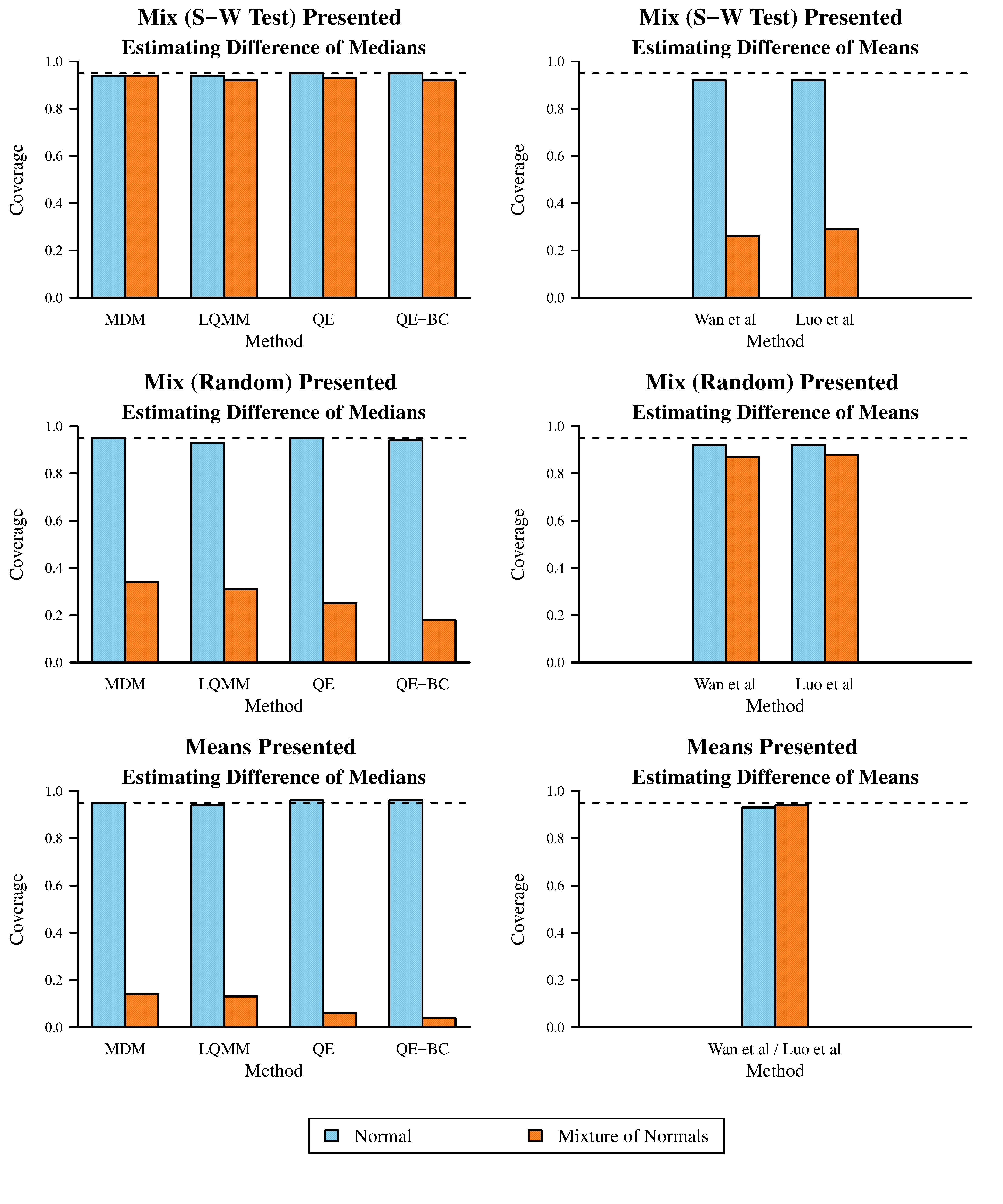}
 \end{figure}

\begin{figure} [H] 
   \centering
    \caption{Bias of the estimate of heterogeneity of the methods by the outcome distribution and summary measures reported in the sensitivity analysis. The methods were applied to the simulated data when the number of studies was 10, the median number of subjects per study was 50, the value of $I^2$ was $25\%$, and the added effect size in the normal distribution case was moderate. \label{ae.sensitivity}}
 \includegraphics[scale=0.9]{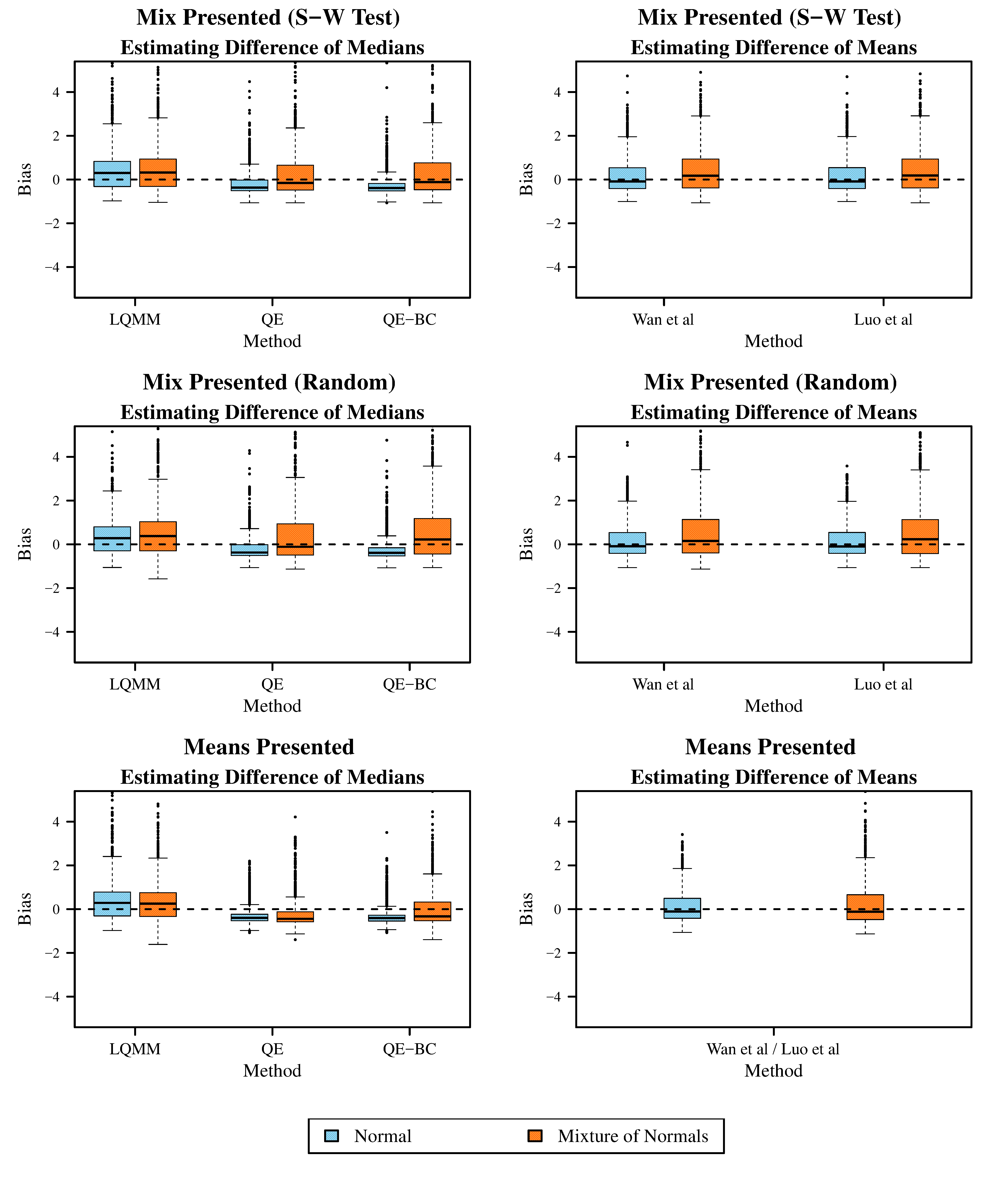}
 \end{figure}

In this section, we provide the figures corresponding to the sensitivity analysis. Figures \ref{re.sensitivity}, \ref{cov.sensitivity}, and \ref{ae.sensitivity} give the relative error of the pooled estimates, coverage of the 95\% CIs of the pooled estimates, and the bias of the heterogeneity estimates, respectively. 

\clearpage

\clearpage 

\clearpage 

\clearpage
\section{Fixed Effect Methods} \label{fixed effect}

In this section, we apply the inverse variance methods in fixed effect analyses. This includes both transformation-based methods as well as the QE and QE-BC methods. Table \ref{table.c} gives the relative error, variance, and coverage of the 95\% CIs of the inverse variances methods with fixed effect pooling.  The results are similar to those of the random effects analyses, and the same overall conclusions and recommendations hold in the fixed effect case.

\clearpage

\begin{table}[t]
\caption {Performance of inverse-variance methods in fixed effect analyses. Performance is measured by relative error, variance, and coverage of the 95\% CIs of the pooled estimate. In the ``Relative Error" columns, the data are presented in the format: median (first quartile, third quartile). The methods were applied to the simulated data when the number of studies was 10, the median number of subjects per study was 50, the value of $I^2$ was $25\%$, and the added effect size in the normal distribution case was moderate.} \label{table.c}
\begin{center}
\begin{tabular}{@{\extracolsep{6pt}}llllllll@{}}
  \hline
  & & \multicolumn{2}{c}{Relative Error} &  \multicolumn{2}{c}{Variance}  & \multicolumn{2}{c}{Coverage}   \\ 
  \cline{3-4} \cline{5-6} \cline{7-8}
  Scenario & Method &  Normal & Mixture &  Normal & Mixture & Normal  & Mixture \\ 
  \hline
 $S_1$ & Wan et al & -2.28 (-13.02, 9.76) & -32.22 (-38.37, -26.49) & 0.32 & 0.29 & 0.82 & 0.04 \\
 & Luo et al  & 0.18 (-10.37, 12.53) & -15.40 (-21.59, -8.55) & 0.29 & 0.32 & 0.81 & 0.40 \\
  & QE  & 0.13 (-10.52, 12.62) & 0.83 (-8.01, 9.34) & 0.33 & 0.29 & 0.91 & 0.95 \\ 
  & QE-BC & -0.10 (-10.33, 11.80) & 1.05 (-7.96, 9.84) & 0.31 & 0.28 & 0.90 & 0.89 \\ 
   $S_2$ & Wan et al & -0.15 (-9.84, 10.30) & -21.73 (-27.52, -16.67) & 0.26 & 0.27 & 0.83 & 0.10 \\ 
  & Luo et al  & -0.16 (-9.96, 10.49) & -21.16 (-27.00, -15.95) & 0.26 & 0.27 & 0.84 & 0.11 \\ 
  & QE & -0.14 (-10.02, 11.22) & 0.18 (-9.16, 8.83) & 0.31 & 0.30 & 0.88 & 0.88 \\ 
  & QE-BC & -0.10 (-10.33, 11.80) & 1.05 (-7.96, 9.84) & 0.31 & 0.28 & 0.90 & 0.89 \\
  $S_3$ & Wan et al & 0.02 (-10.81, 11.57) & 11.55 (4.81, 18.67) & 0.29 & 0.41 & 0.80 & 0.51 \\ 
  & Luo et al  & -0.17 (-9.69, 10.30) & -11.88 (-17.59, -6.65) & 0.24 & 0.27 & 0.85 & 0.49 \\ 
  & QE & 0.08 (-10.47, 12.60) & 0.72 (-8.13, 9.24) & 0.32 & 0.29 & 0.91 & 0.95 \\ 
  & QE-BC & -0.10 (-10.33, 11.80) & 1.05 (-7.96, 9.84) & 0.31 & 0.28 & 0.90 & 0.89 \\ 
  \hline
\end{tabular}
\end{center}
\end{table}

\end{document}